\documentclass[12pt]{article}
\pdfoutput=1 
\usepackage{natbib}
\usepackage{amsmath}
\usepackage{amssymb}
\usepackage{graphicx}
\graphicspath{{./fig/}}
\usepackage[final]{showkeys}
\usepackage{color}
\usepackage{hyperref}

\newcommand\fat\boldsymbol
\newcommand\refeq[1]{Eq.~\ref{#1}}

\title{Spatial forecast postprocessing: The Max-and-Smooth approach}
\author{Stefan Siegert\footnote{Corresponding author, Email: \texttt{s.siegert@exeter.ac.uk}},\\ Ben Hooper, Joshua Lovegrove, Tyler Thomson,\\ Birgir Hrafnkelsson}
\date{Draft version: \today}

\begin{document}
\maketitle

\section*{Abstract}

Numerical weather forecasts can exhibit systematic errors due to simplifying model assumptions and computational approximations.
Statistical postprocessing is a statistical approach to correcting such biases.
A statistical postprocessing model takes input data from a numerical forecast model, and outputs a parametric predictive distribution of a real-world observation, with model parameters learned from past forecast-observation pairs.
In this paper we develop and discuss methods for postprocessing of gridded data.
We show that estimates of postprocessing parameters on a spatial grid can be improved by Bayesian hierarchical modelling with spatial priors.
We use the ``Max-and-Smooth'' approach \citep{hrafnkelsson2021maxandsmooth} to approximate a fully Bayesian inference in two steps. 
First we calculate maximum-likelihood estimates (MLEs) of postprocessing parameters at individual grid points. 
Second we smooth the MLEs using a measurement error model with a spatial prior.
Our approach provides the theoretical basis for the parameter smoothing approach by \citet{kharin2017postprocessing}, and simplifies and generalises the Bayesian hierarchical modelling approach by \citet{moller2015spatially}.
A new derivation of Max-and-Smooth is presented. 
The method is applicable to arbitrary postprocessing models, as illustrated on Model Output Statistics, Logistic Regression, and Nonhomogeneous Gaussian Regression.
We report consistent improvements in forecast accuracy, calibration, and probabilistic skill in postprocessing of temperature and precipitation forecasts.


\section{Introduction}

Forecasts generated by numerical simulators are prone to systematic biases that can be corrected in a postprocessing step \citep{glahn1972use, vannitsem2018statistical}.
A common example is to subtract a constant bias from a temperature forecast that is known to be systematically too warm.
Statistical postprocessing techniques use a data set of past forecasts, and their known verifying observations, to fit a parametric statistical model that takes a forecast as input and outputs a point estimate or a probability distribution for the corresponding future observation.
Due to the dynamic nature of the forecast model and underlying physical system, the statistical behavior of forecast errors can change over space and time. 
Postprocessing models should be able to adapt to such local differences, e.g., by using different statistical models in different locations.
The research presented in this paper addresses the problem of efficiently using spatial forecast and observation data to fit postprocessing models with spatially adaptive parameters.

We differentiate between spatial postprocessing and local postprocessing.
Local postprocessing approaches fit statistical postprocessing models independently at each location, using only data from that location, without accounting for any spatial correlations of data or model parameters.
In spatial postprocessing statistical models are used that can account for spatial correlations; see \citet{schefzik2018ensemble} for a review.
Spatial postprocessing can be applied either at the data level, or the parameter level, or both.
Spatial postprocessing at the data level, i.e., applied at the level of the predicted meteorological variable, aims to produce spatially coherent meteorological fields.
Data-level methods include the Schaake shuffle \citep{clark2004schaake}, ensemble copula coupling \citep{schefzik2013uncertainty}, and Gaussian process models \citep{berrocal2008probabilistic}.
Spatial postprocessing at the parameter level aims to improve estimates of local postprocessing parameters by exploiting similarity of parameters at nearby locations.
Approaches include postprocessing of spatially aggregated data \citep{van2020influence}, spatial smoothing of postprocessing parameters \citep{kharin2017postprocessing, lovegrove2022}, or Bayesian hierarchical modelling with a spatially correlated prior on postprocessing parameters \citep{moller2015spatially}.
Postprocessing at parameter and data level can be combined to improve both, estimates of local model parameter and the spatial coherence of output fields \citep{moller2015spatially}.

This paper is concerned with spatial postprocessing at the parameter level. 
We present a simple method improve local estimates of postprocessing parameters by borrowing strength from data at nearby locations.
Formally the method yields the same results as the Max-and-Smooth approach by \citet{hrafnkelsson2021maxandsmooth}, but the derivation of the method differs slightly from the original proof.
The argument starts from a Bayesian hierarchical model for the meteorological observations, with a spatially correlated prior distribution used for the postprocessing parameters. 
In that way our method is similar to the approach by \citet{moller2015spatially}. 
We show that the posterior distribution of postprocessing parameters can be approximated by first calculating their local maximum likelihood estimates (MLEs), and secondly, spatially smoothing the MLEs. 
In that way our method is similar to the parameter smoothing method proposed by \citet{kharin2017postprocessing}.
The ``Max-and-Smooth'' approach at the same time unifies the approaches of \citet{moller2015spatially} and \citet{kharin2017postprocessing}, it simplifies and speeds up the Bayesian inference used by \citet{moller2015spatially}, and provides a more solid mathematical foundation for the ad-hoc smoothing approach of \citet{kharin2017postprocessing}.
In our method, the degree of spatial smoothing is partly controlled by the estimation uncertainty of the local MLEs, which allows us to learn the smoothing hyperparameters from the data, which is another advantage compared to the approach of \citet{kharin2017postprocessing}.

Temperature and precipitation data used to illustrate our methods are detailed in Sec.~\ref{sec:data}. 
We illustrate the Max-and-Smooth approach on postprocessing parameters from Model Output Statistics (MOS), Logistic Regression (LR), and Nonhomogeneous Gaussian Regression (NGR), which are briefly summarised in Sec.~\ref{sec:ppmethods}.
The main result of the paper is the Bayesian Hierarchical modelling approach to spatial postprocessing and its formulation as a spatial smoothing applied to Maximum likelihood estimates of postprocessing parameters, which is detailed in Sec.~\ref{sec:maxandsmooth}.
Further methodological details are summarised in Sec.~\ref{sec:rw2d} on choice of the spatial prior, in Sec.~\ref{sec:hyperpars} on inference of hyperparameters, Sec.~\ref{sec:rinla} on the R-INLA software package, and Sec.~\ref{sec:eval} on evaluation methods and metrics used to assess the performance of the proposed method.
Sec.~\ref{sec:results} summarises performance improvements due to spatial postprocessing of temperature and precipitation forecasts with LR, MOS and NGR, and Sec.~\ref{sec:discussion} concludes with a summary and discussion.

\section{Data and Methods}
\label{sec:datamethods}

\subsection{Data}
\label{sec:data}

To illustrate and compare the methods we use hindcast data of different meteorological variables and at different forecast lead times. 
Forecast data were downloaded from the ECMWF S2S data base \citep{vitart2017subseasonal}.
Forecasts were generated by ECMWF's seasonal forecast model which has 10 ensemble members perturbed from 1 control forecast. 
Reanalysis data from the ERA5 reanalysis project \citep{hersbach2020era5} were used as pseudo-observations.
All forecast and reanalysis data are available on a $1.5^\circ$ latitude-longitude grid on the rectangular domain between longitudes $15^\circ W$ and $30^\circ E$ and latitudes $36^\circ N$ and $69^\circ N$.
The API requests to download reanalysis and reforecast data are given in the Supplementary Material.

We use 2-metre temperature ensemble forecasts to illustrate and compare performance metrics when postprocessing real-valued quantities.
We use exceedance of precipitation thresholds to illustrate the methods applied to postprocessing probabilities of binary events.
All forecasts were initialised at 00:00UTC on 14 April each year from 2002 to 2021, and forecast lead times 1 day up to 10 days.

\subsection{Postprocessing methods}
\label{sec:ppmethods}

We assume hindcast data at a given lead time are available as ensemble forecasts with $K$ members on a regular latitude-longitude grid with grid point indexed by $s = 1, \dots, S$, and for verification date indexed by $t = 1, \dots, T$. 
Since all analyses are carried out independently for each forecast lead time, the lead time index is suppressed.
We denote the $k$-th ensemble member at grid point $s$ and verification date $t$ by $f_{k,s,t}$, and summarise the ensemble by the ensemble mean 
$$m_{s,t} = \frac1K \sum_{k=1}^K f_{k,s,t}$$
and the ensemble variance 
$$v_{s,t} = \frac{1}{K-1}\sum_{k=1}^K (f_{k,s,t} - m_{s,t})^2.$$

Model output statistics \citep[MOS,][]{glahn1972use} was one of the earliest
postprocessing methods proposed to correct systematic biases of numerical
models, and to predict variables that are not explicitly simulated by the numerical model. 
MOS is essentially a least-squares linear regression, where the verifying observation $y_{s,t}$ is modelled as a linear transformation of the ensemble mean forecast $m_{s,t}$, plus an independent normally distributed error with zero mean and time-constant variance:
\begin{equation}
y_{s,t} \sim N(\mu^{MOS}_{s,t}, (\sigma^{MOS}_s)^2)
\end{equation}
where 
\begin{align}
\mu^{MOS}_{s,t} & = \alpha^{MOS}_{s} + \beta^{MOS}_{s} (m_{s,t} - \bar{m}_s), \\
\log\left(\sigma^{MOS}_s\right)^2 & = \tau^{MOS}_s,
\end{align}
and where $\bar{m}_s = \frac1T \sum_{t=1}^T m_{s,t}$ is the time-averaged ensemble mean.
We parameterise the model in terms of the log-variance $\tau^{MOS}_s$ to have a real-valued parameter for which we can use a normal distribution as a spatial prior for Bayesian inference.
Centering the covariates by subtracting $\bar{m}_s$ diagonalises the information matrix (below) and hence makes MOS parameter estimates asymptotically independent.

Let $\fat{y}_s$ denote the vector of observed data at location $s$. 
The log-likelihood function of the MOS model is given by
\begin{equation}
\begin{aligned}[b]
& \log p(\fat{y}_s | \fat\theta^{MOS}_s) = \log p(\fat{y}_s|\alpha^{MOS}_s, \beta^{MOS}_s, \tau^{MOS}_s) \\
& = -\frac{T}{2}\log2\pi - \frac{T}{2} \tau^{MOS}_s - \frac{1}{2} e^{-\tau^{MOS}_s} \sum_{t=1}^{T} \left\{ y_{s,t} - (\alpha^{MOS}_s + \beta^{MOS}_s (m_{s,t} - \bar{m}_s)) \right\}^2.
\end{aligned}
\end{equation}
The maximum likelihood estimators (MLEs) at grid point $s$ are then given by
\begin{equation}
\begin{aligned}
\hat\alpha^{MOS}_s & = \bar{y}_s = \frac{1}{T}\sum_{t=1}^{T} y_{s,t},\\
\hat\beta^{MOS}_s & = \frac{\sum_{t=1}^{T} (y_{s,t} - \bar{y}_s)(m_{s,t}-\bar{m}_s)}{\sum_{t=1}^{T}(m_{s,t} - \bar{m}_s)^2}\text{, and}\\
\hat\tau^{MOS}_s & = \log \left\{ \frac{1}{T} \sum_{t=1}^{T} \left[ y_{s,t} - (\hat\alpha^{MOS}_s + \hat\beta^{MOS}_s (m_{s,t}-\bar{m}_s)) \right]^2 \right\}.
\end{aligned}
\end{equation}
We estimate the precision of the sampling distribution of the MLEs by the observed information matrix \citep{efron1978assessing} which is given by 
\begin{equation}
\hat{J}_s = - \frac{\partial^2 \log p(\fat{y}_s | \fat{\theta}_s)}{\partial \fat{\theta}_s \partial \fat{\theta}_s'}\Bigg|_{\fat\theta_s = \hat{\fat\theta}_s} = \begin{pmatrix} Te^{-\hat\tau_s} & 0 & 0 \\ 0 & \sum_{t=1}^{T} (m_{s,t}-\bar{m}_t)^2 e^{-\hat\tau_s} & 0 \\ 0 & 0 & \frac{T}{2} \end{pmatrix}.
\end{equation}

Logistic regression \cite[][ch.~7]{wilks2011statistical} can be used to calculate
probability of precipitation from deterministic model output of total precipitation amounts. 
We use simple logistic regression to predict the event that the daily precipitation amount exceeds $2.5$mm, using the ensemble mean total precipitation as a single covariate. 
Let the binary prediction target $y_{s,t}$ be equal to $1$ if precipitation amount at location $s$ and verification time $t$ exceeds $2.5$mm, and $y_{s,t}=0$ otherwise.
Define the forecast probability $p^{LR}_{s,t} = P(y_{s,t} = 1)$, which is calculated from the ensemble mean precipitation amount forecast $m_{s,t}$ by 
\begin{equation}
p^{LR}_{s,t} = \left(1+e^{-(\alpha^{LR}_s + \beta^{LR}_s m_{s,t})}\right)^{-1}.
\label{eq:logreg}
\end{equation}
With $\fat\theta^{LR}_s = (\alpha^{LR}_s, \beta^{LR}_s)$, the log-likelihood is given by
\begin{equation}
\begin{aligned}[b]
& \log p(\fat{y}_s|\fat{\theta}^{LR}_s) \\
& = - \sum_{t=1}^T \left[ (1-y_{s,t})(\alpha^{LR}_s + \beta^{LR}_s m_{s,t}) + \log\left(1 + e^{-(\alpha^{LR}_s + \beta^{LR}_s m_{s,t})}\right)\right]
\end{aligned}
\end{equation}
There are no closed form solutions for the MLEs $\hat{\fat\theta}^{LR}_s = (\hat\alpha_s^{LR}, \hat\beta^{LR}_s)$, so they are obtained by numerical optimsation using the \texttt{R} function \texttt{optim}. 
The observed information of the logistic regression model at location $s$ is given by
\begin{equation}
\hat{J}_s = \sum_{t=1}^T \hat{p}_{s,t}^{LR}(1-\hat{p}_{s,t}^{LR}) \begin{pmatrix}1 & m_{s,t}\\m_{s,t} & m_{s,t}^2\end{pmatrix}
\end{equation}
where $\hat{p}_{s,t}^{LR}$ is Eq.~\ref{eq:logreg} evaluated at $\hat{\fat\theta}_s^{LR}$.

Non-homogeneous Gaussian regression \citep[NGR, ][]{gneiting2005calibrated} is similar to MOS but allows for non-constant forecast variances. 
NGR accounts for the fact that in addition to the ensemble mean, the ensemble variance can be informative for the forecast distribution.
In NGR the real-valued observations are normally distributed
\begin{equation}
y_{s,t} \sim N(\mu^{NGR}_{s,t}, (\sigma^{NGR}_{s,t})^2)
\label{eq:ngr-lik}
\end{equation}
where the forecast mean $\mu^{NGR}_{s,t}$ depends linearly on the the ensemble mean $m_{s,t}$ (as in MOS), and the forecast variance $(\sigma^{NGR}_{s,t})^2$ depends linearly on the ensemble variance $v_{s,t}$ through
\begin{equation}
\begin{aligned}
\mu^{NGR}_{s,t} & = \alpha_s + \beta_s m_{s,t} \\
\sigma^{NGR}_{s,t} & = (e^{\gamma_s} + e^{\delta_s} v_{s,t})^{1/2}.
\label{eq:ngr-par}
\end{aligned}
\end{equation}
The log-likelihood follows from Eq.~\ref{eq:ngr-lik} and Eq.~\ref{eq:ngr-par}.
The MLEs $\hat{\fat{\theta}}^{NGR}_s = (\hat\alpha^{NGR}_s, \hat\beta^{NGR}_s, \hat\gamma^{NGR}_s, \hat\delta^{NGR}_s)'$ are not available in closed form, and so we obtain them via numerical optimisation.
The observed information can be calculated analytically, but the expressions are complicated, and we found that numerical approximations (returned by \texttt{optim}) are accurate. 
Note that our implementation of NGR differs from that of \cite{gneiting2005calibrated}, who used minimum CRPS estimation rather than MLE, and parameterised the variance as $\gamma + \delta^2 v_{s,t}$.

We are working with samples of $T=20$ when fitting local MLEs of postprocessing parameters.
Such small sample sizes can lead to poor convergence and ``exploding'' parameter estimates in numerical optimisation.
To help the numerical optimiser and somewhat regularise the parameter estimates we added $-0.5 \times 10^{-4}[(\alpha^{LR})^2 + (\beta^{LR})^2]$ to the LR log-likelihood, and, similarly $-0.5 \times 10^{-4}[(\alpha^{NGR})^2 + (\beta^{NGR})^2 + (\gamma^{NGR})^2 + (\delta^{NGR})^2]$ to the NGR log-likelihood during numerical optimisation.
This additive term can be interpreted as an $L_2$ (ridge) penalty that shrinks parameter estimates towards zero \citep{hastie2009elements}, or equivalently as independent zero-mean normal prior distributions with standard deviation 100 applied to all parameters.
While we will refer to parameter estimates as MLEs, strictly speaking they should be called PMLEs (penalised MLEs) or MAP (maximum a-posteriori) estimates.
The effect of the penalty term is negligible in cases where we optimiser converges without the penalty, and it leads to successful convergence to ``reasonable'' parameter values when the optimiser fails to converge without the penalty.

\subsection{Bayesian hierarchical modelling and Max-and-Smooth}
\label{sec:maxandsmooth}

All our postprocessing methods are such that observations $y_{s,t}$ at spatial locations $s=1,\dots,S$ and time instances $t=1,\dots,T$ are independently distributed with some distribution $p(y_{s,t}|\fat\theta_s)$ that is conditional on spatially varying (but time-constant) postprocessing parameters $\fat\theta_s = (\alpha_s, \beta_s,\dots)'$. The dependency of $p(y_{s,t}|\fat\theta_s)$ on a vector of covariates $\fat{z}_{s,t}$ is understood, and not written out explicitly. 
Let observations at all locations and times be collected in the vector $\fat{y}$ and all postprocessing parameters collected in the vector $\fat\theta$.
Conditional independence then implies
\begin{equation}
\log p(\fat{y}|\fat\theta) = \sum_{s=1}^S \sum_{t=1}^T \log p(y_{s,t} | \fat\theta_s).
\end{equation}
For Bayesian inference of $\fat\theta$ we assume $\fat\theta$ to have a prior normal distribution with mean vector $\fat\mu_\theta$ and precision matrix $Q_\theta$, i.e., the prior log-density of $\fat\theta$ can be written as
\begin{equation}
\log p(\fat\theta) = const - \frac12 (\fat\theta - \fat\mu_\theta)'Q_\theta(\fat\theta - \fat\mu_\theta),
\label{eq:prior}
\end{equation}
where $const$ denotes a generic additive constant that does not depend on $\fat\theta$.
Defining
\begin{equation}
f_s(\fat\theta_s) = \sum_{t=1}^T \log p(y_{s,t} | \fat\theta_s),
\end{equation}
the posterior log-density of $\fat\theta$ is given by
\begin{equation}
\log p(\fat\theta|\fat{y}) = const + \sum_{s=1}^S f_s(\fat\theta_s) - \frac12 (\fat\theta - \fat\mu_\theta)'Q_\theta (\fat\theta - \fat\mu_\theta).
\end{equation}

To make posterior inference of $\fat\theta$ analytically tractable, we Taylor-approximate $f_s(\fat\theta_s)$ to second order around its mode $\hat{\fat\theta}_s$
\begin{equation}
f_s(\fat\theta_s) \approx \tilde{f}_s(\fat\theta_s) = const - \frac12 (\fat\theta_s - \hat{\fat\theta}_s)'\hat{J}_s(\fat\theta - \hat{\fat\theta}_s)
\label{eq:laplaceapprox}
\end{equation}
where $\hat{\fat\theta}_s = \mathrm{argmax} f_s(\fat\theta_s)$ is the vector of maximum likelihood estimates (MLEs) at spatial location $s$, and $\hat{J}_s$ is the observed information matrix at location $s$ evaluated at the local MLEs $\hat{\fat\theta}_s$.
Denote the combined vector of MLEs at all spatial locations by $\hat{\fat\theta}$, and the combined observed information matrix $\hat{J} = J(\hat{\fat\theta}) = -\frac{\partial^2}{\partial \fat\theta \partial \fat\theta'} \sum_{s=1}^S f_s(\fat\theta_s) |_{\hat{\fat\theta}}$.
If the elements of the combined parameter vector $\fat\theta$ are ordered as $\fat\theta' = (\fat\theta_1', \dots, \fat\theta_S')$, then the combined information matrix is block-diagonal as $\hat{J} = \mathrm{bdiag}(\hat{J}_1, \dots, \hat{J}_S)$.
Under the second-order Taylor approximation of $f_s(\fat\theta_s)$ we have an approximation of the posterior distribution of $\fat\theta$ given by
\begin{equation}
\begin{aligned}[c]
\log p(\fat\theta | \fat{y}) \approx & \log \tilde{p}(\fat\theta | \fat{y})\\
& = const -\frac12 (\fat\theta - \hat{\fat\theta})' \hat{J} (\fat\theta - \hat{\fat\theta}) - \frac12 (\fat\theta - \fat\mu_\theta)' Q_\theta (\fat\theta - \fat\mu_\theta)\\
& = const - \frac12 \fat\theta'(Q_\theta + \hat{J})\fat\theta + (Q_\theta \fat\mu_\theta + \hat{J} \hat{\fat\theta})'\fat\theta
\label{eq:posttheta}
\end{aligned}
\end{equation}
which is the log-density of a multivariate normal distribution in canonical form \citep{rue2005gaussian}, and hence the posterior distribution of $\fat\theta$ is approximately multivariate normal with conditional expectation
\begin{equation}
E(\fat\theta|\fat{y}) = (Q_\theta + \hat{J})^{-1}(Q_\theta \fat\mu_\theta + \hat{J} \hat{\fat\theta})
\label{eq:pmean}
\end{equation}
and conditional variance matrix
\begin{equation}
Var(\fat\theta | \fat{y}) = (Q_\theta + \hat{J})^{-1}.
\label{eq:pvar}
\end{equation}
The Gaussian approximation of the posterior is the central approximation underlying the Integrated Nested Laplace Approximations method \citep[INLA,][]{rue2009approximate} used by \citet{moller2015spatially} for spatial inference of postprocessing parameters in a Bayesian hierarchical model.
Their spatial prior is a stochastic partial differential equation (SPDE) which corresponds to a specific choice of the prior precision matrix $Q_\theta$.
Our spatial prior and corresponding choice of $Q_\theta$ is introduced in Section \ref{sec:rw2d}.


We now present an alternative derivation of Eq.~\ref{eq:posttheta} which makes the connection between the approximate Bayesian hierarchical modelling approach with R-INLA used by \citet{moller2015spatially}, and the parameter smoothing approach for spatial postprocessing proposed by \citet{kharin2017postprocessing}.
According to asymptotic likelihood theory \citep{schervish2012theory} the MLE $\hat{\fat\theta}_s$ has an asymptotic normal (sampling) distribution centered on the ``true'' parameter value $\fat\theta_s$ and with asymptotic precision given by the expected information $I_s = -E\left[\frac{\partial^2}{\partial \fat\theta_s \partial \fat\theta_s'}\log p(\fat{y}_s|\fat\theta_s)\right]$.
Since the observed information $\hat{J}_s$ is often a more appropriate measure of the precision of $\hat{\fat\theta}_s$ at estimating $\fat\theta_s$ \citep{efron1978assessing} we may approximate the sampling distribution by
\begin{equation}
\hat{\fat\theta}_s \overset{\cdot}{\sim} N(\fat\theta_s, \hat{J}_s^{-1}).
\label{eq:mledist}
\end{equation}
We interpret the MLEs $\hat{\fat\theta}_s$ as ``noisy measurements'' of the ``true'' parameters $\fat\theta_s$, with normally distributed measurement error with zero mean and variance $\hat{J}_s^{-1}$. 
Assuming that the MLEs at different locations are independent we thus have
\begin{equation}
\log p(\hat{\fat\theta} | \fat\theta) = const - \frac12 (\hat{\fat\theta} - \fat\theta)'\hat{J} (\hat{\fat\theta} - \fat\theta).
\end{equation}
To infer $\fat\theta$ from the observed values of $\hat{\fat\theta}$ we specify the same multivariate normal prior for $\fat\theta$ as in \refeq{eq:prior}. 
Then we have for the posterior of $\fat\theta$
\begin{align}
\log p(\fat\theta|\hat{\fat\theta}) & =  const + \log p(\hat{\fat\theta}|\fat\theta) + \log p(\fat\theta)\\
& = const - \frac12 \fat\theta' (Q_\theta + \hat{J})\fat\theta + (Q_\theta \fat\mu_\theta + \hat{J}\hat{\fat\theta})'\fat\theta
\end{align}
which implies
\begin{equation}
\begin{aligned}
E(\fat\theta | \hat{\fat\theta}) & = (Q_\theta + \hat{J})^{-1} (Q_\theta \fat\mu_\theta + \hat{J}\hat{\fat\theta})\text{, and}\label{eq:posttheta-ms}\\
Var(\fat\theta | \hat{\fat\theta}) & = (Q_\theta + \hat{J})^{-1}.
\end{aligned}
\end{equation}
The approximate posterior of $\fat\theta$ given $\hat{\fat\theta}$ is identical to the approximate posterior of $\fat\theta$ given $\fat{y}$ in Eq.~\ref{eq:posttheta}.
We have thus shown that a normal approximation of the likelihood function in a Bayesian hierarchical model with a multivariate normal prior results in the same posterior as a model that assumes that the MLEs $\hat{\fat\theta}$ are ``noisy measurements'' of the ``true'' parameters $\fat\theta$, with error (co-)variances derived from their asymptotic sampling distribution.
Since Eq.~\ref{eq:mledist} models the MLEs as ``truth plus error'', the inferred posterior expectations of $\fat\theta$ will usually appear spatially smoother than the MLEs $\hat{\fat\theta}$. 
The inference of the posterior expectation of $\fat\theta$ can thus be regarded as a smoothing applied to the local MLEs, which is also what \citet{kharin2017postprocessing} proposed to use spatial information more effectively in postprocessing of gridded data.

In summary, to approximate spatial postprocessing by a Bayesian hierarchical model we can proceed in a two-steps:
First we postprocess forecasts locally, by calculating the MLEs and the observed information matrix at each location separately.
Secondly we ``post-postprocess'' the MLEs by spatially smoothing them, under a measurement error model using a spatially correlated prior and measurement error (co-)variances derived from the observed information.
\citet{hrafnkelsson2021maxandsmooth} hence dubbed that approximation ``Max-and-Smooth''.
While yielding identical results, our derivation of Max-and-Smooth differs slightly from the original. 
Here, we motivate the measurement error model via the asymptotic sampling distribution of the MLEs. 
\citet{hrafnkelsson2021maxandsmooth} simply ``reverse-engineered'' the measurement error model from the Laplace approximation of the posterior, and showed that it yields identical results.

\subsection{The RW2D spatial prior}
\label{sec:rw2d}

The 2-dimensional random walk \citep[RW2D,][]{rue2005gaussian} is a first order
intrinsic autoregressive model on a lattice indexed by row $i=1,\dots,N_i$ and column $j=1,\dots,N_j$.
An RW2D random field $W \in \mathbb{R}^{N_i \times N_j}$ is defined recursively by assuming an independent normal distribution of 2-dimensional increments,
\begin{equation}
W_{i,j} - \frac14 \left( W_{i-1,j} + W_{i+1,j} + W_{i, j-1} + W_{i,j+1} \right) \sim N\left(0,\frac{1}{\kappa}\right).
\end{equation}
The relationship between $W_{i,j}$ and the average over its nearest neighbors induces spatial correlation. 
The higher the precision parameter $\kappa$ the closer $W_{i,j}$ is (on average) to its neighbors, and hence $\kappa$ controls the smoothness of the random field $W$.
For more background on the RW2D see \citet{besag2005first}, who discuss the
relationship between intrinsic autoregressions and the de Wijs process, and
their examples include the RW2D as a special case.

Let $W \in \mathbb{R}^{N_i \times N_j}$ be a realisation of an RW2D on a $N_i \times N_j$ grid, and $E \in \mathbb{R}^{N_i \times N_j}$ a matrix of independent standard normally distributed random variates. 
Then the RW2D can be written as the Sylvester equation
\begin{equation}
D_{N_i} W + W D_{N_j} = \kappa^{-1/2} E,
\label{eq:rw2d-mat}
\end{equation}
where $D_n$ denotes the tri-diagonal $n\times n$ matrix
\begin{equation}
D_n = \frac14 \begin{pmatrix} 1 & -1 \\ -1 & 2 & -1 \\  & \ddots & \ddots & \ddots \\ && -1 & 2 & -1 \\ &&& -1 & 1\end{pmatrix}.
\end{equation}
Let the $N_iN_j$-vectors $\fat{w}$ and $\fat{e}$ be the vectorisations of $W$ and $E$ obtained by column-stacking. 
Then Eq.~\ref{eq:rw2d-mat} can be written as
\begin{equation}
\kappa^{1/2} D\fat{w}  = \fat{e}
\end{equation}
with 
\begin{equation}
D = D_{N_i} \otimes I_{N_j} + I_{N_i} \otimes D_{N_j},
\end{equation}
where $I_n$ denotes the $n\times n$ identity matrix \citep{shores2007applied}.
It follows that $\fat{w}$ is a normally distributed random vector with mean zero and precision matrix
\begin{equation}
Q_{RW2D} = \kappa R_{RW2D} = \kappa D'D
\label{eq:Qrw2d}
\end{equation}
\citep{rue2005gaussian}. 
We will use $Q_{RW2D}$ as the spatial prior precision $Q_{\theta}$ to infer postprocessing parameters throughout this paper.

The $S \times S$ matrix $Q_{RW2D}$ (where $S = N_iN_j$) is rank-deficient with $\mathrm{rank}(Q_{RW2D}) = S - 1$, and hence $Q_{RW2D}$ does not have a well-defined inverse. 
Marginal variances and covariances of the RW2D process are thus ill-defined, and numerical sampling from the RW2D process problematic. 
In our application, $Q_{RW2D}$ is used merely as a prior to infer the posterior mean and variance of $\fat\theta$ (Eq.~\ref{eq:pmean} and Eq.~\ref{eq:pvar}). 
These posterior quantities depend on inverses of $(Q_{RW2D} + \hat{J})$ which is usually of full rank, and hence the rank deficiency of $Q_{RW2D}$ is benign.

For joint spatial inference of multiple postprocessing parameters (e.g. $\alpha^{LR}$ and $\beta^{LR}$ in logistic regression), we specify independent spatial RW2D priors for them. 
If the elements of the full parameter vector $\fat\theta$ are ordered as $\fat\theta = (\alpha_1, \dots, \alpha_S, \beta_1, \dots, \beta_S, \dots)'$, and each of $\fat\alpha, \fat\beta, \dots$ have independent RW2D priors, then the joint prior for $\fat\theta$ is a multivariate normal distribution with mean zero and block-diagonal precision matrix
\begin{equation} 
Q_\theta = \begin{pmatrix}Q_\alpha & &\\& Q_\beta & \\&& \ddots\end{pmatrix} = \begin{pmatrix}\kappa_\alpha R_{Rw2D} & &\\& \kappa_\beta R_{RW2D} & \\&& \ddots\end{pmatrix}.
\label{eq:Qthetablockdiag}
\end{equation} 
The precision parameters $\fat\kappa = (\kappa_\alpha, \kappa_\beta, \dots)'$ are hyperparameters.

The matrices $Q_{RW2D}$ and $\hat{J}$ are sparse matrices. 
Sparsity allows for efficient numerical solutions of linear systems such as Eq.~\ref{eq:pmean}, and efficient random sampling which can be used to calculate approximate marginal variances from Eq.~\ref{eq:pvar} \citep{rue2005gaussian}.
In this paper, we use sparse matrix methods implemented in the R package \texttt{Matrix} \citep{matrixpackage}.

\subsection{Inferring hyperparameters}
\label{sec:hyperpars}

For spatial inference of postprocessing parameters with a fixed prior precision matrix $Q_\theta$ we have to specify the precision parameter $\kappa$ for each parameter, which controls the amount of smoothing when inferring $\fat\theta$ from $\hat{\fat\theta}$.
Let the vectors $\fat\kappa$ and $\fat\theta$ denote vectors of, respectively, hyperparameters and postprocessing parameters.
We momentarily write the spatial prior for $\fat\theta$ as conditional on $\fat\kappa$, i.e. $p(\fat\theta|\fat\kappa)$ and specify a prior distribution $p(\fat\kappa)$.
The posterior log-density of the hyperparameters $\fat\kappa$, given the MLEs $\hat{\fat\theta}$, is given by
\begin{equation}
\log p(\fat\kappa | \hat{\fat\theta}) = \log p(\hat{\fat\theta} | \fat\theta) + \log p(\fat\theta|\fat\kappa) + \log p(\fat\kappa) - \log p(\fat\theta | \hat{\fat\theta}, \fat\kappa) - \log p(\hat{\fat\theta})
\label{eq:postkappa}
\end{equation}
where $p(\hat{\fat\theta}|\fat\theta)$ follows from the normally distributed measurement error model of the MLEs, and the posterior $p(\fat\theta | \hat{\fat\theta}, \fat\kappa)$ is approximated by a normal distribution as in Eq.~\ref{eq:posttheta-ms}. 
The term $p(\hat{\fat\theta})$ is a constant that does not depend on $\fat\kappa$, and the value of $\fat\theta$ when evaluating the rhs of Eq.~\ref{eq:postkappa} can be set arbitrarily.

For the analyses in this paper we infer $\fat\theta$ by empirical Bayes, i.e. setting the precision hyperparameters $\fat\kappa = (\kappa_\alpha, \kappa_\beta, \dots)'$ to the fixed values that maximise Eq.~\ref{eq:postkappa} with respect to $\fat\kappa$. 
Ignoring any terms that do not depend on $\fat\kappa$, we thus estimate $\fat\kappa$ by the point estimator
\begin{equation}
\begin{aligned}
\hat{\fat\kappa} & = \underset{\fat\kappa}{\mathrm{argmax}} \Big[-\frac12 (\hat{\fat\theta} - \fat\theta^*)'\hat{J}(\hat{\fat\theta} - \fat\theta^*) + \frac12 \log \det Q_\theta\\
& \quad \qquad \qquad - \frac12 {\fat\theta^*}' Q_\theta \fat\theta^* + \log p(\fat\kappa) - \frac12 \log \det (Q_\theta + \hat{J})\Big].
\end{aligned}
\label{eq:kappahat}
\end{equation}
where $\fat\theta^* = E(\fat\theta | \hat{\fat\theta}, \fat\kappa)$ given by Eq.~\ref{eq:posttheta-ms}.
Since $Q_\theta$ is block-diagonal (Eq.~\ref{eq:Qthetablockdiag}) and $R_{RW2D}$ has rank $S-1$ and is independent of $\fat\kappa$, the log-determinant of $Q_\theta$ is given by $\log\det Q_\theta = const + (S - 1)(\log \kappa_\alpha + \log \kappa_\beta + \dots)$ \citep{rue2005gaussian}.
We use independent Exponential prior distributions with rate $5\times 10^{-5}$ for the elements of $\fat\kappa$, hence 
\begin{equation}
\log p(\fat\kappa) = const - 5\times 10^{-5} (\kappa_\alpha + \kappa_\beta + \dots).
\end{equation}
To find $\hat{\fat\kappa}$ we use numerical optimisation in the \texttt{R} function \texttt{optim} with default settings.

The obtained value of $\hat{\fat\kappa}$ is plugged into $Q_\theta$ to calculate the smoothed postprocessing via Eq.~\ref{eq:pmean}. 
Our empirical Bayes approach ignores the estimation uncertainty due to inferring $\fat\kappa$.
To assess the effect of this simplification we compare our results to parameter estimates obtained with R-INLA (cf. Sec.~\ref{sec:rinla}) which approximates a fully Bayesian inference where $\fat\kappa$ is integrated out rather than fixed.

\subsection{R-INLA}
\label{sec:rinla}

The use of multivariate normal priors with sparse precision matrices, and the normal approximation of the likelihood we used in Eq.~\ref{eq:laplaceapprox} are also the main ingredients used in the R-INLA package \citep{rinla} for large scale Bayesian inference.
R-INLA has the option of inferring the posterior of the latent field $\fat\theta$ by empirical Bayes, i.e. $p(\fat\theta|\fat{y}) = p(\fat\theta | \fat{y}, \hat{\fat\kappa})$ or by numerically integrating over the posterior of hyperparameters, i.e., solving $p(\fat\theta | \fat{y}) = \int p(\fat\theta | \fat{y}, \fat\kappa) p(\fat\kappa|\fat{y}) d\fat\kappa$.
There are additional refinements to the various approximations as detailed in \citet{rue2009approximate}, but essentially our approach is very similar to R-INLA. 
A review of R-INLA's spatial modelling capabilities can be found in \citet{bakka2018spatial}, and coding tutorials and examples are available in \url{https://www.r-inla.org/examples-tutorials}\footnote{Last accessed 6 July 2022.}, \citet{blangiardo2015spatial} and \citet{gomez2020bayesian}.

\citet{moller2015spatially} use R-INLA to infer $p(\fat\theta|\fat{y})$, by specifying a Bayesian hierarchical model for the meteorological observations $\fat{y}$ directly, without the intermediate step of calculating MLEs first.
While it is possible to specify RW2D and many other spatial priors in R-INLA, it cannot define all likelihood functions commonly used for forecast postprocessing.
For example, while Bayesian linear regression is readily implemented in R-INLA, it is not possible\footnote{as of version 22.01.12} to specify a spatial prior for the variance parameter, or make the variance dependent on covariates as in NGR. 
R-INLA is thus not applicable in many spatial post-processing applications.

Bayesian Logistic Regression with spatial priors is, however, implemented in R-INLA, so we can use R-INLA as a reference implementation.
We can also assess the effects of our empirical Bayes approach of fixing hyperparameters at $\hat{\fat\kappa}$ compared to integrating them out.
R-INLA includes corrections to improve the normal approximation of the likelihood function, whose benefits we will also be able to assess in a comparison of spatial logistic regression with R-INLA.

Under an additional approximation of the observation matrix, R-INLA can be used to perform the smoothing step of Max-and-Smooth.
Consider the diagonal approximation of $\tilde{J}$ for $\hat{J}$
\begin{equation}
\tilde{J} = \left[\mathrm{diag}(\hat{J}^{-1})\right]^{-1},
\label{eq:diag-approx}
\end{equation}
i.e. the variance matrix of the MLEs is a diagonal matrix with elements equal to their marginal asymptotic variances and covariances between MLEs are ignored.
In the case of MOS that approximation is exact because the MLEs are asymptotically independent.
When $Q_\theta$ is block-diagonal, with one block per parameter, and the information matrix is diagonal (either by construction as in MOS, or by approximation), then the posterior mean $(Q_\theta + \tilde{J})^{-1}\tilde{J}\hat{\fat\theta}$ can be decomposed into updates of individual parameters, i.e. $E(\fat\alpha | \hat{\fat\alpha})= (Q_\alpha + \tilde{J}_\alpha)^{-1} \tilde{J}_\alpha \hat{\fat\alpha}$, etc, where $Q_\alpha$ is the RW2D prior precision for $\fat\alpha$.
The diagonal approximation allows smoothing the MLEs individually and independently, which is straightforward with R-INLA.
The approximation Eq.~\ref{eq:diag-approx} was used in \citet{johannesson2022approximate} for model selection, and \citet{lovegrove2022} used it to apply INLA for spatial postprocessing with MOS.
Example \texttt{R} code for smoothing a spatial field of MLEs with known sampling variances with R-INLA is provided in the Supplementary Material.

\subsection{Model evaluation}
\label{sec:eval}

We illustrate and discuss the performance of the proposed methods by postprocessing the forecast data presented in Section \ref{sec:data}.
In all analyses we first apply local postprocessing by maximum likelihood estimation separately at each grid point and lead time. 
We apply spatial smoothing to the 2-dimensional fields of local MLEs using Max-and-Smooth with independent RW2D priors.
The quality of postprocessing models is evaluated and compared by deterministic and probabilistic verification metrics.

We are working with complete data on a rectangular grid throughout.
There are a total of $S$ grid points, indexed by the single spatial index $s = 1,\dots,S$.
At each grid point $s$ and lead time $\tau$, a number of $T$ forecast-observation pairs is available at verification times indexed by $t=1,\dots, T$.
Depending on context, we average verification metrics either over both, spatial coordinate $s$ and verification time $t$, or, to highlight spatial differences of performance, only over verification time $t$.
Verification scores are always calculated separately at each lead time $\tau$, and hence the dependency on lead time $\tau$ is suppressed in mathematical notation.

The accuracy of postprocessed temperature forecasts (i.e.~, expectations of the postprocessed predictive distributions) are evaluated by calculating the mean squared errors (MSE) \citep{deque2012deterministic}. 
Let $\mu_{s,t}$ denote the mean of the forecast distribution at spatial location $s$ and verification time $t$, and let $y_{s,t}$ be the corresponding verifying observation.
The MSE is the space-time averaged squared prediction error
\begin{equation}
MSE = \frac{1}{ST} \sum_{s =1}^S \sum_{t=1}^T (y_{s,t} - \mu_{s,t})^2.
\end{equation}

Forecast probabilities $p_{s,t}$ of precipitation exceedances (binary yes/no, encoded as $y_{s,t}=1$ or $0$, respectively) are evaluated by the Brier score \citep{brier1950verification}.
The average Brier score is given by the mean squared difference between forecast probabilities $p_{s,t} = P(y_{s,t} = 1)$ and binary event indicators
\begin{equation}
BS = \frac{1}{ST} \sum_{s=1}^S \sum_{t=1}^T (y_{s,t} - p_{s,t})^2.
\end{equation}
The Brier score is a proper scoring rule that rewards truthful representations of the probability distribution of the target event frequency \cite{broecker2012probability}. 
Like the MSE it is negatively oriented, assigning lower values to better forecasts.

Probabilistic temperature forecast, issued as full probability density functions (pdfs) over all possible outcomes are evaluated by the logarithmic score \citep[also Logscore or Ignorance,][]{bernardo1979expected, roulston2002evaluating}.
Let $f_{s,t}(x)$ denote the forecast pdf at grid point $s$ and verification time $t$, and $y_{s,t}$ the real-valued verifying observation.
The Logscore is given by
\begin{equation}
LS = \frac{1}{ST}\sum_{s=1}^{S}\sum_{t=1}^T \left[-\log f_{s,t}(y_{s,t})\right].
\end{equation}
The Logscore is proper and negatively oriented.

Our goal is to assess whether, and how much, parameter smoothing can improve forecasts of future observations in operational settings.
To assess out-of-sample performance, and avoid results that are biased by overfitting, we apply leave-one-out cross validation \citep{hastie2009elements}. 
Specifically, we leave each individual verification time $t=1,\dots,T$ out in turn, estimate parameters (unsmoothed and smoothed) and hyperparameters on the data set comprising the remaining $T-1$ verification times, and calculate evaluation metrics by predicting observations at the left-out verification time $t$.
We repeat this process $T$ times and average evaluation metrics over all $T$.
In our case, the relatively small number of verification times ($T=20$), as well as the efficiency of the spatial smoothing method makes this ``brute-force'' leave-one-out cross validation practically feasible, without having to resort to approximate model selection critiera \citep{hastie2009elements, piironen2017comparison}.

\section{Results}
\label{sec:results}

In this section we apply Max-and-Smooth to spatial postprocessing of precipitation forecasts with Logistic Regression, and spatial postprocessing of temperature forecasts with Model Output Statistics (MOS) and Nonhomogeneous Gaussian Regression (NGR).
The focus of the analyses is on the comparison of out-of-sample accuracy and probabilistic skill of forecasts before and after spatial smoothing of post-processing parameters.

\subsection{Logistic regression}
\label{sec:logreg}

We fit Logistic Regression to estimate the probability of threshold exceedance based on the total precipitation ensemble mean.
We estimate postprocessing parameters by maximum likelihood estimation as described in Sec.~\ref{sec:ppmethods}.
We then calculate posterior means of postprocessing parameters via Max-and-Smooth, by applying Eq.~\ref{eq:posttheta-ms}, with hyperparameters estimated using Eq.~\ref{eq:kappahat}.
We also use the R-INLA package to fit postprocessing parameters in a Bayesian logistic regression model with spatial RW2D priors on the intercept and slope parameters. 
The main difference between R-INLA and our Max-and-Smooth implementation is that hyperparameters are integrated out, rather than plugged in.
We also evaluate the local climatological forecasts (the unconditional mean rate of threshold exceedances) as a benchmark.

\begin{figure}
\begin{minipage}{.45\textwidth}
\includegraphics{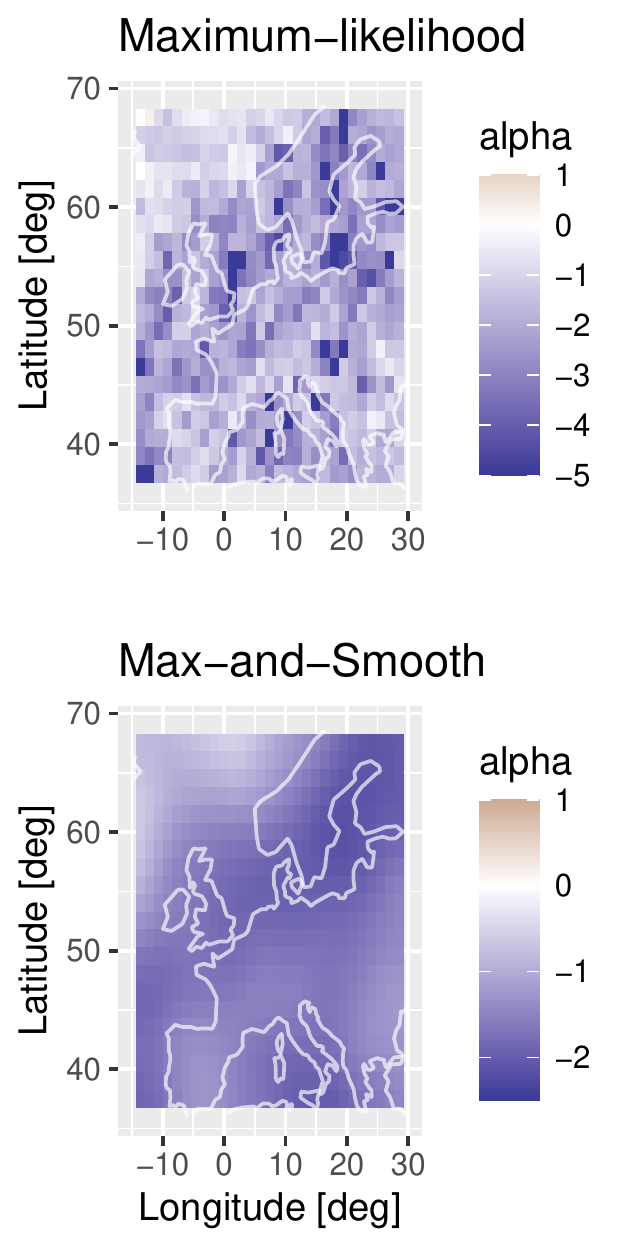}
\end{minipage}
\begin{minipage}{.45\textwidth}
\includegraphics{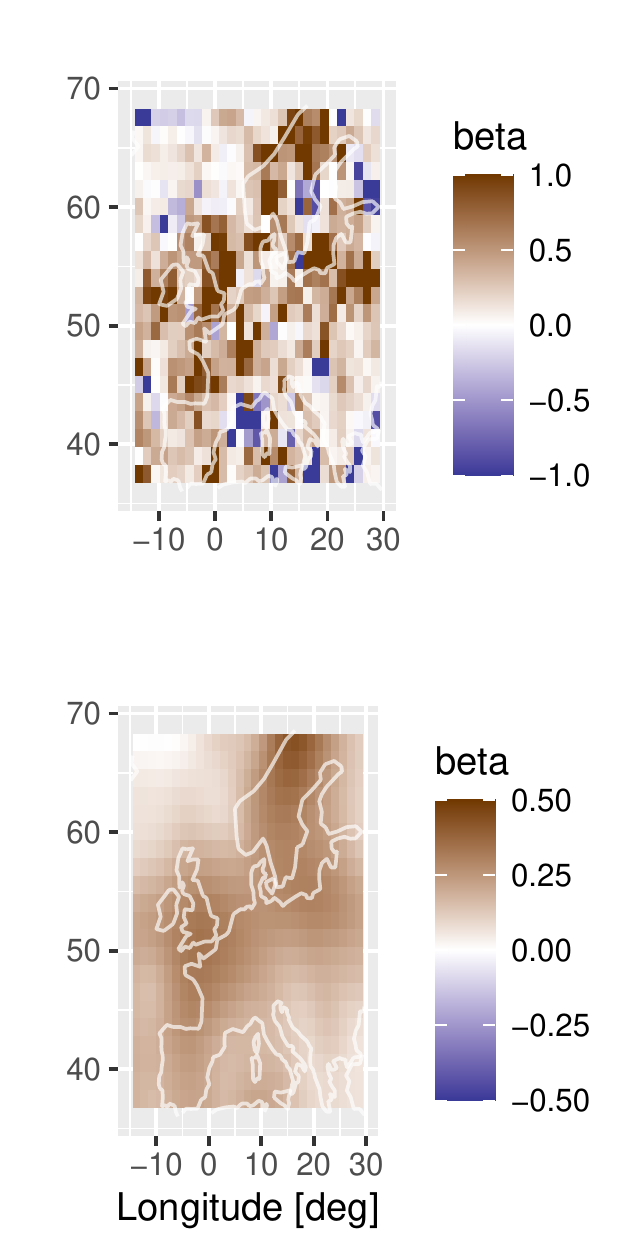}
\end{minipage}
\caption{Parameter estimates of a logistic regression of the probability of exceeding precipitation of $2.5$mm at 1 day lead time. Upper panels: Local maximum-likelihood estimates of regression parameters. Lower Panel: Posterior means obtained by Max-and-Smooth.}
\label{fig:logreg-alphabeta}
\end{figure}

Figure \ref{fig:logreg-alphabeta} compares local MLEs with posterior means of postprocessing parameters calculated with Max-and-Smooth.
The MLEs fluctuate a lot spatially, and so Bayesian inference with a spatially correlated prior has a strong smoothing effect on the parameter estimates. 
After applying the smoothing step, only large scale patterns of variability remain. 
A noteable result is that after spatial smoothing all slope parameters of the regression are strictly positive.

\begin{figure}
\begin{center}
\includegraphics{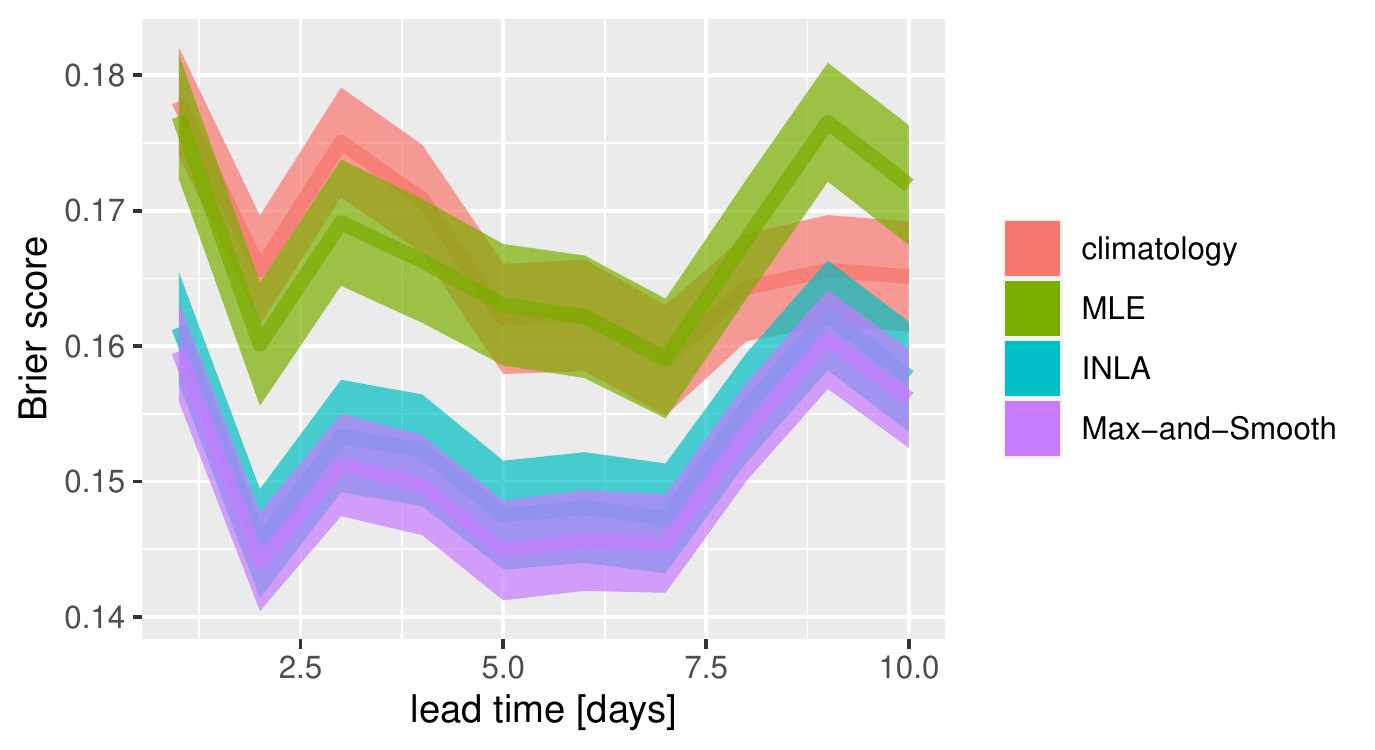}
\caption{Brier scores (mean $\pm$ two standard deviations) of climatological forecast, logistic regression using local MLEs, INLA, and Max-and-Smooth.}
\label{fig:logreg-brier}
\end{center}
\end{figure}

Figure \ref{fig:logreg-brier} shows the spatially averaged leave-one-out Brier scores obtained by local postprocessing, and obtained by spatial post-processing using Max-and-Smooth and R-INLA.
We note from Figure \ref{fig:logreg-brier} that predicting the threshold exceedance based on the ensemble mean seems to be a difficult forecasting problem with generally low skill.
This is indicated by the poor performance of the local MLE logistic regression compared to climatology, even at short lead times.
However, forecast skill can be improved substantially by spatial postprocessing.
The absolute Brier Score improvement of Max-and-Smooth over MLE is between $0.14$ and $0.18$, which corresponds to relative improvements between $8.5\%$ and $11.1\%$.
We further note that Max-and-Smooth achieves similar forecast skill as INLA, which suggests that integrating out hyperparameters has little effect compared to fixing $\fat\kappa$ at a point estimate.

\subsection{MOS}
\label{sec:mos}

In this section we study the skill of temperature forecasts postprocessed with MOS, and compare the performance of postprocessing parameters estimated with MLE and Max-and-Smooth.

\begin{figure}
\begin{center}
\includegraphics{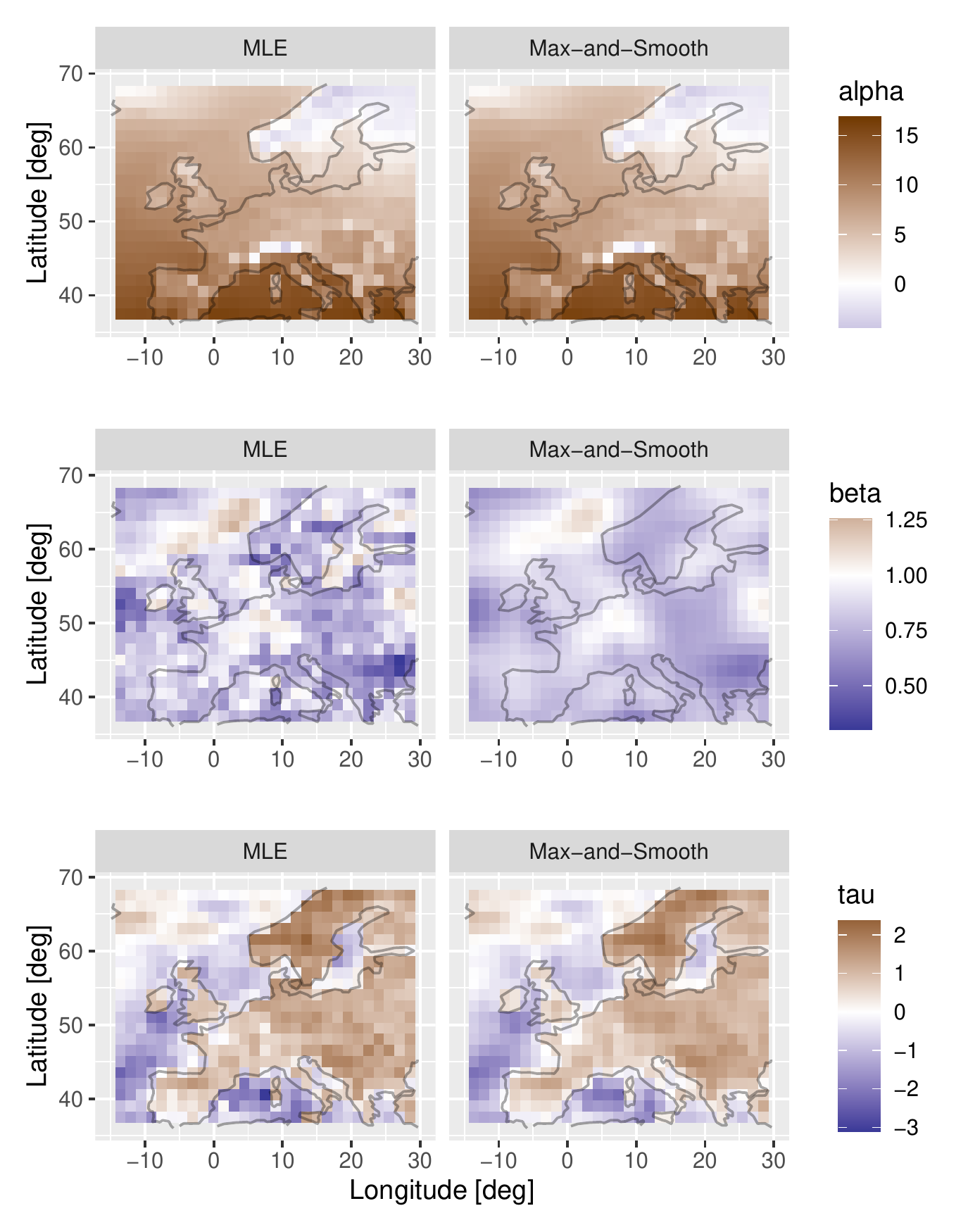}
\end{center}
\caption{Spatial maps of MOS coefficients before and after smoothing. The forecast lead time is 1 day.}
\label{fig:mos-coef-maps}
\end{figure}

Figure \ref{fig:mos-coef-maps} show MOS parameter estimates obtained by MLE and Max-and-Smooth.
In all parameters most of the spatial variations (small and large scale) are preserved after smoothing. 
Hence the postprocessed forecasts obtained by MLE and Max-and-Smooth will not differ much, and we can only expect small differences in forecast skill, if any.
The slope parameters undergo more smoothing than the intercept and log-variance parameters.

\begin{figure}
\begin{center}
\includegraphics{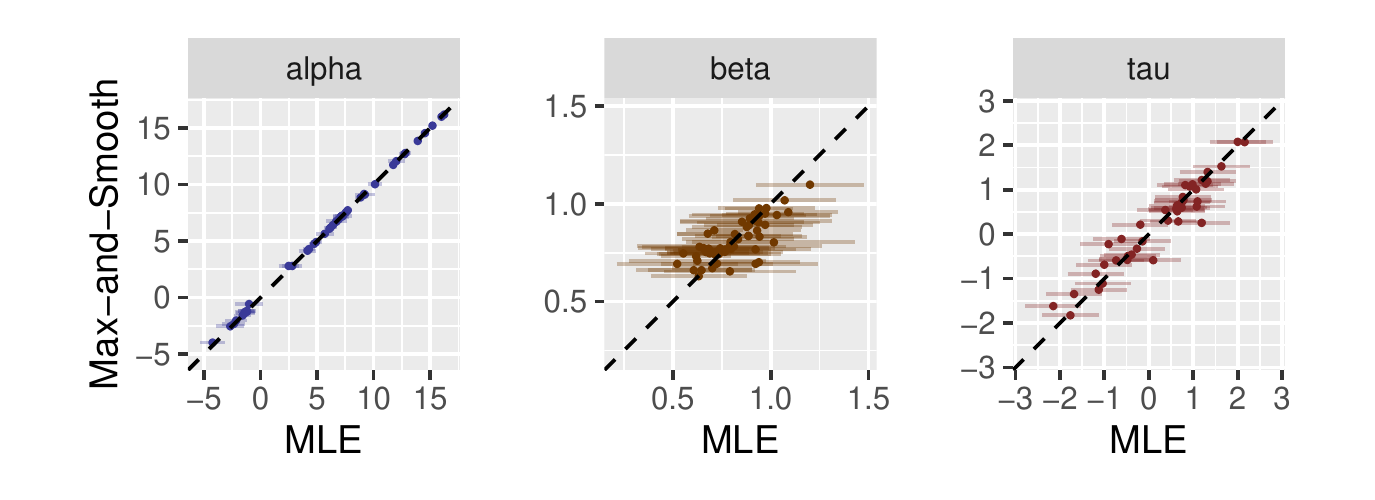}
\caption{Scatter plots of (50 randomly sampled) Max-and-Smooth parameter estimates over the corresponding MLEs. The half-widths of the horizontal lines are equal to two standard deviations of the asymptotic sampling distribution of the MLEs.}
\label{fig:mos-coef-scatter}
\end{center}
\end{figure}

The scatter plots of MOS coefficients in Figure \ref{fig:mos-coef-scatter} explain why the slopes are smoothed more than the intercepts and log-variances.
The asymptotic errors of the intercept MLEs are small, and hence the inferred value of $\alpha_s$ at each location is highly constrained to be close to $\hat\alpha_s$, resulting in very little variation during smoothing. 
On the other hand, the asymptotic errors of the slopes are relatively large. 
Hence $\hat\beta_s$ is less informative about $\beta_s$, and the spatial model has more freedom to move the parameters around, resulting in more smoothing.

\begin{figure}
\begin{center}
\includegraphics{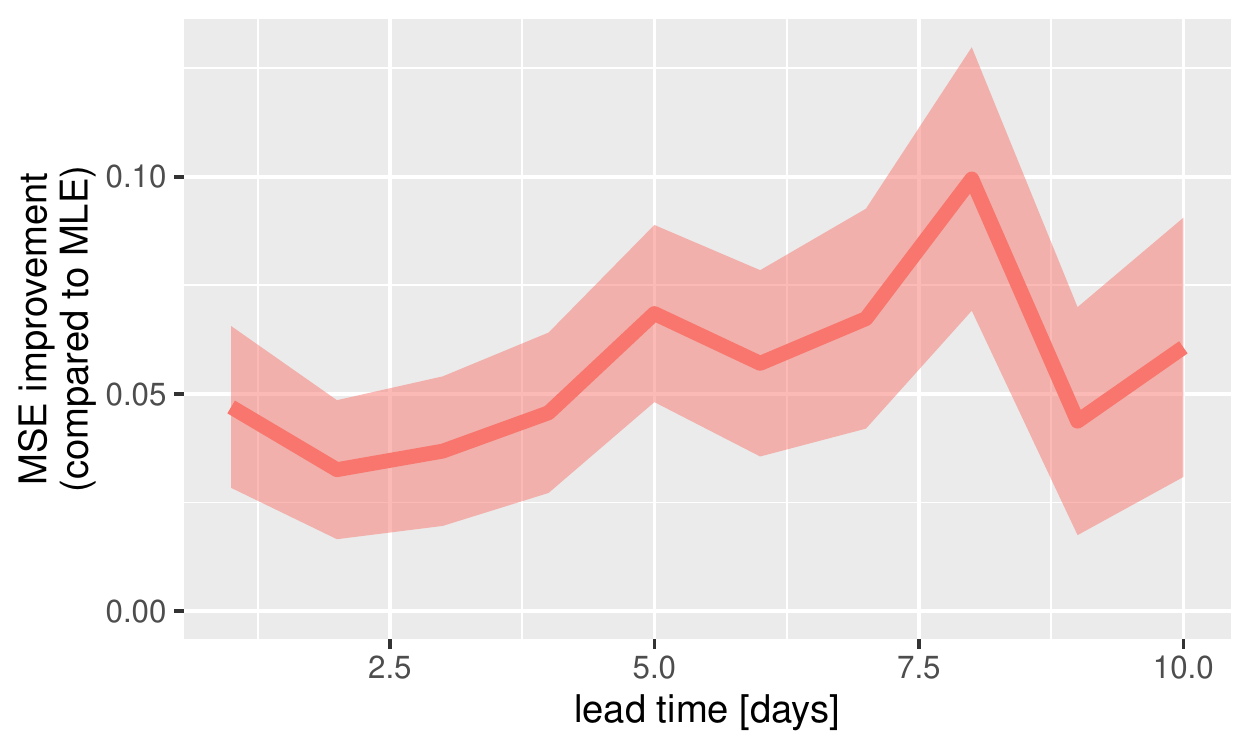}
\caption{Mean squared prediction error difference ($\pm$ 2 standard deviations) between postprocessed forecasts using local MLE vs. using Max-and-Smooth. Spatial parameter smoothing with Max-and-Smooth improves the MSE at all lead times.}
\label{fig:mos-msediff}
\end{center}
\end{figure}

Figure \ref{fig:mos-msediff} shows MSE differences between local and spatially postprocessed temperature forecasts.
Consistent with results in section \ref{sec:logreg}, there is a small, but systematic improvement in MSE at all lead times.

\begin{figure}
\begin{center}
\includegraphics{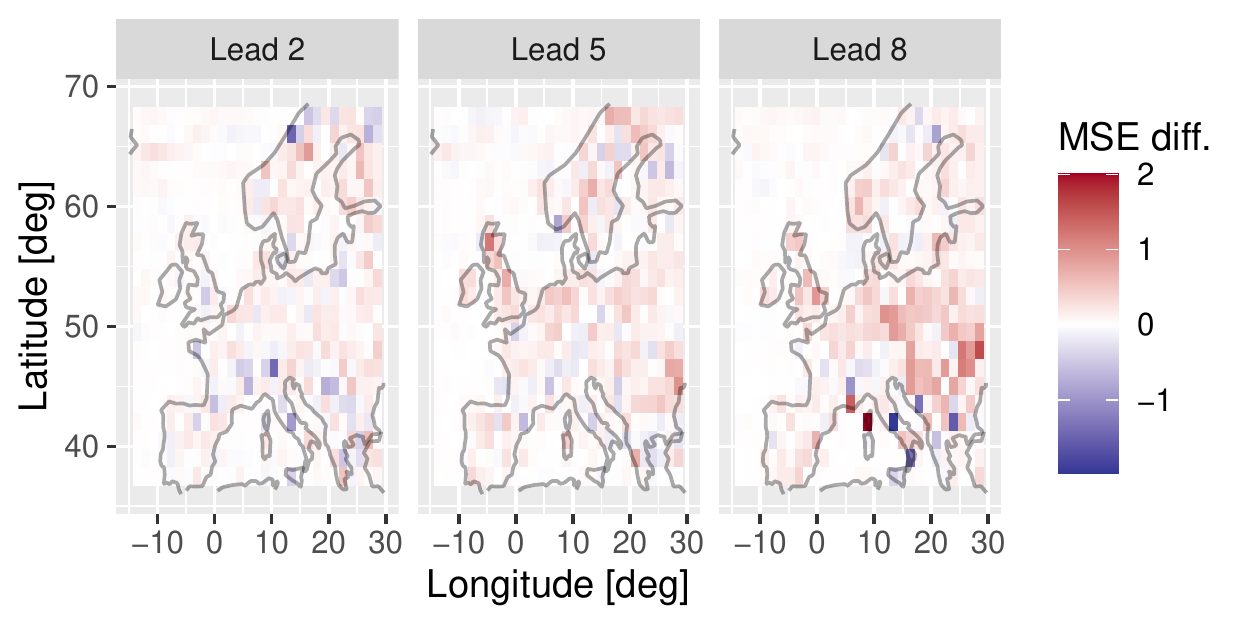}
\caption{Improvement of mean squared prediction error of Max-and-Smooth vs.~MLE.}
\label{fig:mos-msediff-spat}
\end{center}
\end{figure}

Figure \ref{fig:mos-msediff-spat} shows the spatial distribution of MSE differences for lead times 2, 5, and 8 days.
At most locations the MSE differences are positive indicating a consistent improvement of forecast accuracy due to Max-and-Smooth.
Bigger improvements in MSE are seen over land than over sea.
There are a few locations where MSE deteriorates when using Max-and-Smooth.
But the typical magnitudes of positive differences are greater than those of negative differences, resulting in an overall average improvement.

\begin{figure}
\begin{center}
\includegraphics{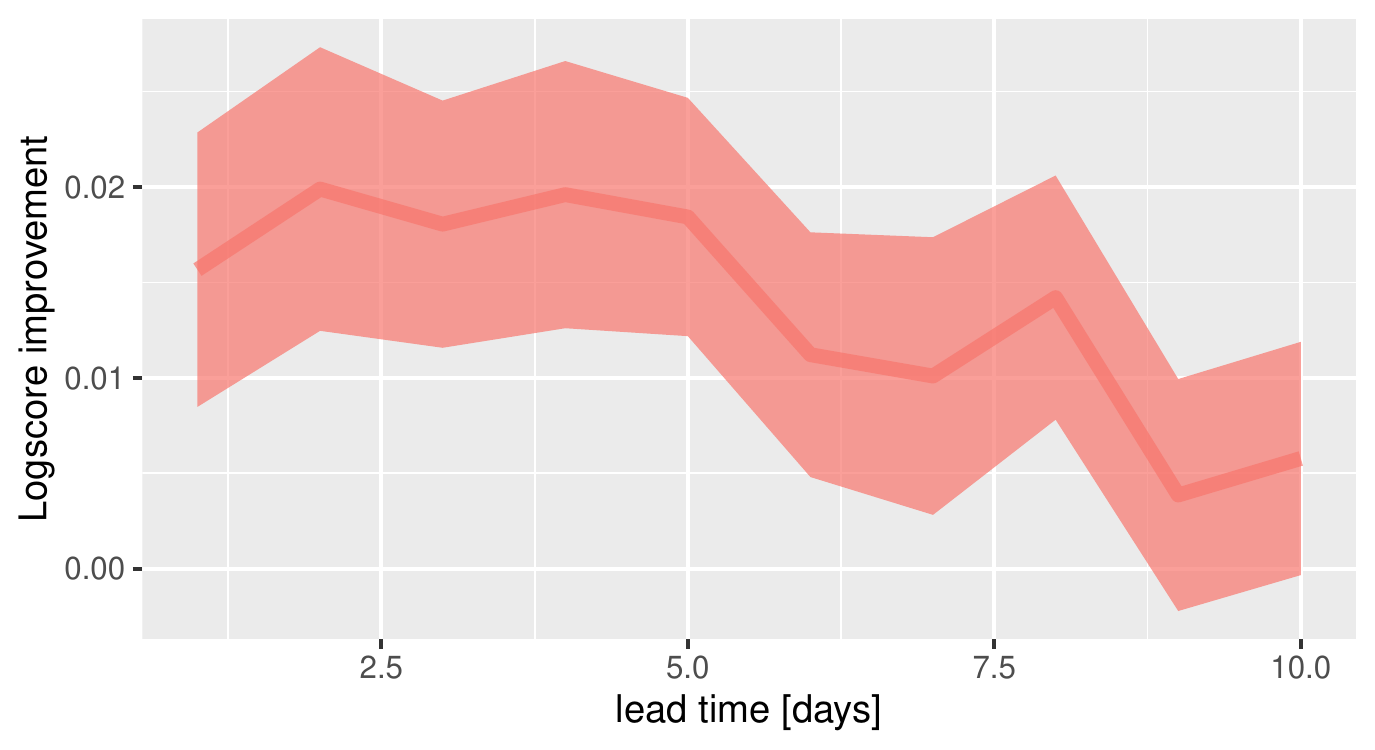}
\caption{Spatially averaged Logscore difference of forecasts postprocessed with MOS, using MLE parameters and Max-and-Smooth parameters. Positive values indicate improvements due to Max-and-Smooth.}
\label{fig:mos-logs}
\end{center}
\end{figure}

Analysis of probabilistic forecast skill by Logscores shown in Figure \ref{fig:mos-logs} confirms the improvement seen in terms of MSE.
The forecast distributions obtained with Max-and-Smooth have better Logscores than forecasts obtained with the unsmoothed MLEs.
The improvement is between $0.004$ and $0.02$, indicating that the Max-and-Smooth distributions assign between $\exp(0.004)\approx 1.004$ and $\exp(0.02)\approx 1.02$ times more density to observations than the MLE distributions.

\subsection{Non-homogeneous Gaussian regression}

Here use instantaneous ensemble means and variances to postprocess temperature forecasts by NGR, as described in section \ref{sec:ppmethods}.
By letting the forecast distribution be informed by the ensemble variance through the parameters $\gamma_s$ and $\delta_s$, we allow narrower (wider) ensembles to generate a narrower (wider) postprocessed forecast distribution.
As before, we first estimate grid-point wise NGR parameters by MLE, and subsequently smooth them using Eq.~\ref{eq:posttheta-ms}, with smoothing parameters $\fat\kappa$ estimated with Eq.~\ref{eq:kappahat}.
We also smooth postprocessing parameters individually with R-INLA, using the diagonal approximation of the information matrix discussed in Sec.~\ref{sec:rinla}.
Smoothing the MLEs independently is only an approximation to ``proper'' Max-and-Smooth because covariances between MLEs are ignored.

\begin{figure}
\begin{center}
\includegraphics{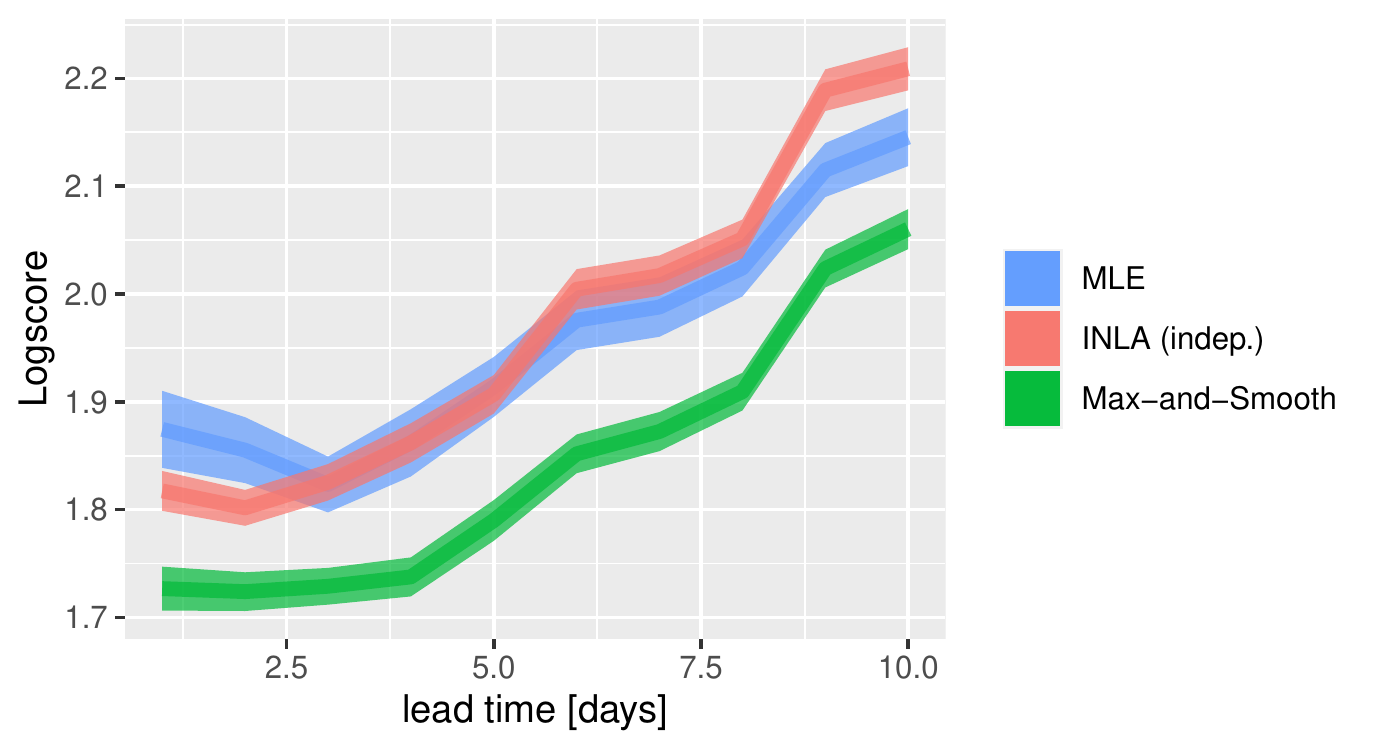}
\caption{Spatially averaged Logscores of NGR models using local MLEs, spatial parameter smoothing with R-INLA assuming independence, and smoothing with Max-and-Smooth accounting for parameter dependence. Smaller values indicate better forecasts.}
\label{fig:ngr-logscore}
\end{center}
\end{figure}

Figure \ref{fig:ngr-logscore} shows average Logscores as function of lead time of local MLE forecasts, forecasts based on individually smoothed postprocessing parameters with R-INLA, and forecasts based on jointly smoothed parameters using Max-and-Smooth. 
Logscores generally increase with increasing forecast lead time, as would be expected. 
Max-and-Smooth consistently outperforms both the local MLE forecasts, and the independently smoothed INLA forecasts, highlighting the benefit of spatial smoothing as well as the importance of the covariances between MLEs.
Interestingly, at lead times above 5 days, the independently smoothed parameters perform worse than the unsmoothed MLEs.
The improvements in Logscores are between $0.09$ (lead time 10 days) and $0.15$ (lead time 1 day), indicating that applying Max-and-Smooth leads to forecast densities that assign between $e^{0.09} \approx 1.09$ and $e^{0.15}\approx 0.16$ more density to the observations than the local MLEs.

We further analysed probabilistic skill by the Continuous Ranked Probability Score \citep[CRPS,][]{matheson1976scoring}, and forecast calibration by Probability Integral Transform (PIT) histograms \citep{gneiting2007probabilistic}. 
Details of these analyses can be seen in the Supplementary Material.
In summary, average CRPS of forecasts obtained with independently smoothed parameters is consistently worse than with unsmoothed parameters, and consistently better when parameters are postprocessed with Max-and-Smooth, accounting for depdendency between parameters. 
PIT histograms obtained with unsmoothed parameters appear U-shaped, but close to uniform after parameter smoothing with either method, indicating considerable improvement of forecast calibration.

\begin{figure}
\begin{center}
\includegraphics{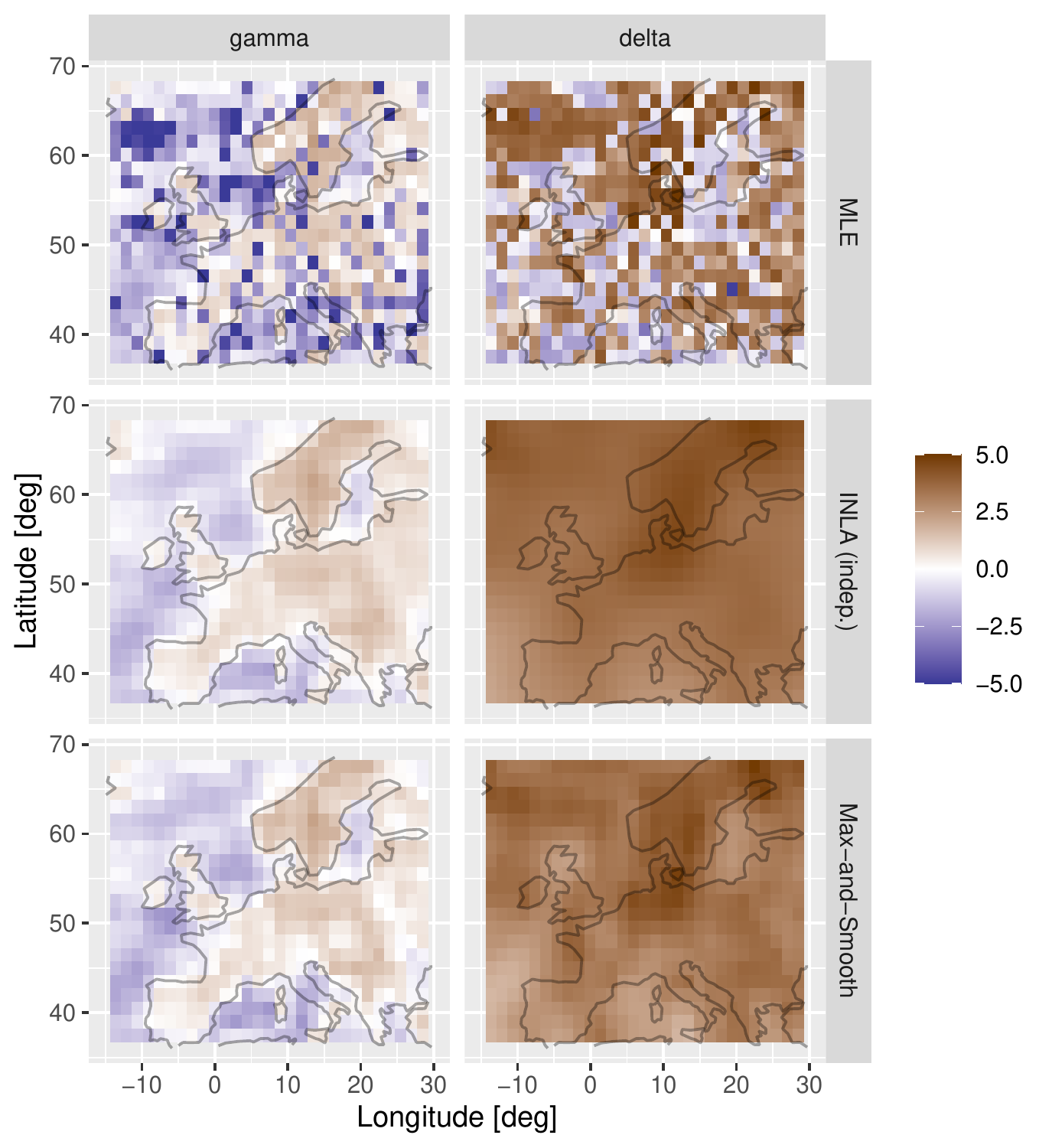}
\caption{Estimates of NGR variance parameters $\gamma$ and $\delta$, lead time 1 day, obtained by MLE, smoothing with R-INLA assuming independence, and Max-and-Smooth taking into account dependence between postprocessing parameters.} 
\label{fig:ngr_coefs}
\end{center}
\end{figure}

Figure \ref{fig:ngr_coefs} shows the NGR variance parameters $\gamma$ and $\delta$ for the 1 day lead time temperature forecasts, estimated by the three methods.
The MLEs appear spatially very noisy, and the effect of spatial smoothing by both methods is large.
The smoothed estimates obtained with R-INLA and obtained with Max-and-Smooth look generally similar, but the Max-and-Smooth estimates retain more spatial structure than the independent R-INLA estimates.
Under the diagonal approximation of the information matrix, a parameter at grid point $i,j$ is adjusted only based on values of the same parameter at neighboring grid points. 
In ``full'' Max-and-Smooth, which takes correlation between parameters into account, the smoothed value of, say, $\delta$ at grid point $i, j$ depends not only on values of $\delta$ at neighboring grid points, but also on nearby values of $\alpha$, $\beta$ and $\gamma$.
That dependency on more information might explain the more detailed spatial structure of Max-and-Smooth estimates compared to the independent R-INLA estimates.
The effect is large enough to lead to the notable improvements of forecast skill seen in Fig.~\ref{fig:ngr-logscore}.
We also note, the smoothed versions of the constant additive variance offset, $e^{\gamma_s}$, are systematically greater than 1 over land, and smaller than 1 over sea, indicating that ensemble variances exhibit more underdispersion over land than over sea.
That pattern is consistent across all lead times (not shown).

\section{Discussion}
\label{sec:discussion}

Spatial smoothing of postprocessing parameters has been previously proposed to use spatial information more effectively \citep{kharin2017postprocessing}.
We have shown that parameter smoothing can be motivated by approximating Bayesian inference of postprocessing parameters in a Bayesian hierarchical model with a spatially correlated prior.
Unlike \citet{kharin2017postprocessing}, our approach explicitly accounts for the estimation uncertainty of the local postprocessing parameters.
MLEs with smaller estimation uncertainty are smoothed less than more uncertain parameters, and so considering estimation uncertainty is useful to constrain the smoothing hyperparameters.

Max-and-Smooth applied to postprocessing leads to consistent improvements of postprocessed forecasts in applications of logistic regression, linear regression (MOS), and nonhomogenous Gaussian regression.
Improvements were demonstrated in terms of accuracy of the forecast mean, as measured by their mean squared errors, probabilistic skill as measured by Brier scores, Logscores, and CRPS, and probabilistic calibration as measured by PIT histograms.
Overall, we deem the evidence that spatial postprocessing is beneficial convincing, and did therefore not carry out any significance testing.

The proposed approach is simple to apply, as it is based on a straightforward linear transformation of local parameter estimates.
As such it operates on results that practitioners generate routinely.
Spatial postprocessing can be regarded as a ``post-postprocessing'' step, hence maintaining a breakdown of postprocessing steps into local postprocessing by calculating MLEs (``Max'' ...) and subsequent spatial postprocessing of the local MLEs (... ``and Smooth'').
We hope that such a clear separation of methodologies will allow for easier adoption of the method in operational settings.

There are several directions for further study and application of the proposed methodology.
Max-and-Smooth involves approximations and simplifications whose impact on reliability, accuracy, and skill of postprocessed forecasts can be studied further. 
We have not implemented a fully Bayesian framework in which parametric uncertainty in hyperparameters and postprocessing parameters is propagated into the predictive distributions.
Such an extension is straightforward and might yield further improvements, although the comparison between INLA and Max-and-Smooth in Sec.~\ref{sec:logreg} suggest that these effects are likely small.
Max-and-Smooth presented here is a spatial postprocessing method at the parameter level, and as such does not produce forecasts that are spatially consistent with observations.
\citet{moller2015spatially} showed that an additional postprocessing step can be applied to independent samples from postprocessed forecast distributions to recover spatial consistency, and such an approach would also be straightforward to apply in our framework.
For a preliminary application of combining an empirical copula approach and Max-and-Smooth to generate spatially coherent Generalised Extreme Value distributions, see \citet{johannesson2022approximate}.
\citet{gneiting2005calibrated} estimated NGR parameters by minimum-CRPS estimation, rather than MLE.
The theory to derive Max-and-Smooth as an approximation of inference in a Bayesian hierarchical model presented here requires use of the MLEs. 
But on a purely pragmatic level, spatial smoothing of minimum-CRPS estimators is certainly possible, especially considering the asymptotic Normality of such estimators \citep{yuen2014crps}.
Lastly, spatial modelling allows for estimation of postprocessing parameters in locations where no observation data are available.
A spatial modelling approach could thus be used to avoid the use of reanalysis data and apply postprocessing directly to ``real observations'', using (spatially sparse) station data.

\section*{Acknowledgments}

In addition to the software packages cited in the article, data analyses and visualisations relied on the R packages \texttt{tidyverse} \citep{wickham2019welcome} and \texttt{Rnaturalearth} \citep{south2017rnaturalearth}.
We thank Tommy Irons and Annette M\"oller for helpful discussions and encouraging feedback.

\bibliographystyle{abbrvnat}
\bibliography{main}

\clearpage

\appendix

\section*{Supplementary material}

\section{CDS API requests to download reanalysis data}

The following python code retrieves the ERA5 temperature reanalysis data via the Copernicus Data Store (CDS) API\footnote{\url{https://cds.climate.copernicus.eu/api-how-to} [Last accessed: 9 August 2022]}.

{\footnotesize
\begin{verbatim}
import cdsapi
c = cdsapi.Client()
c.retrieve(
  'reanalysis-era5-single-levels',
  {
    'product_type': 'reanalysis',
    'variable': '2m_temperature',
    'year': ['2002', '2003', '2004', '2005', '2006', 
             '2007', '2008', '2009', '2010', '2011', 
             '2012', '2013', '2014', '2015', '2016', 
             '2017', '2018', '2019', '2020', '2021'], 
    'month': '04',
    'day': ['15', '16', '17', '18', '19', 
            '20', '21', '22', '23', '24'
    ],
    'time': [
        '00:00', '06:00', '12:00', '18:00'
    ],
    'format': 'grib',
    'area': [70, -15, 35, 30],
  },
  'era5-t2m.grib')
\end{verbatim}
}

To retrieve ERA5 total precipitation reanalysis data, change the request as follows:

{\footnotesize
\begin{verbatim}
c.retrieve(
  ...
  'variable': 'total_precipitation',
  ...
  'time': ['00:00', '01:00', '02:00', '03:00', '04:00', '05:00', 
           '06:00', '07:00', '08:00', '09:00', '10:00', '11:00', 
           '12:00', '13:00', '14:00', '15:00', '16:00', '17:00', 
           '18:00', '19:00', '20:00', '21:00', '22:00', '23:00'],
  ...    
  'era5-tp.grib')
\end{verbatim}
}

\section{MARS API requests to download ensemble forecasts}

The following python code retrieves the 2-metre temperature ensemble forecasts that was used in this paper from the S2S database via the ECMWF Web API\footnote{\url{https://www.ecmwf.int/en/computing/software/ecmwf-web-api} [Last accessed: 9 August 2022]}.

{\footnotesize
\begin{verbatim}
from ecmwfapi import ECMWFDataServer
server = ECMWFDataServer()
server.retrieve({
  "class": "s2",
  "dataset": "s2s",
  "date": "2022-04-14",
  "expver": "prod",
  "hdate": "2002-04-14/2003-04-14/2004-04-14/2005-04-14/
            2006-04-14/2007-04-14/2008-04-14/2009-04-14/
            2010-04-14/2011-04-14/2012-04-14/2013-04-14/
            2014-04-14/2015-04-14/2016-04-14/2017-04-14/
            2018-04-14/2019-04-14/2020-04-14/2021-04-14",
  "levtype": "sfc",
  "model": "glob",
  "number": "1/2/3/4/5/6/7/8/9/10",
  "origin": "ecmf",
  "param": "167",
  "step": "0-24/24-48/48-72/72-96/96-120/120-144/
           144-168/168-192/192-216/216-240",
  "stream": "enfh",
  "time": "00:00:00",
  "type": "pf",
  "target": "reforecast-t2m.grib"
})
\end{verbatim}
}

To retrieve the precipitation forecasts used in this paper, replace the following parameters

{\footnotesize
\begin{verbatim}
server.retrieve({
  ...
  "param": "228228",
  "step": "24/48/72/96/120/144/168/192/216/240",
  "target": "reforecast-tp.grib"
  ...
})
\end{verbatim}
}

\section{Max-and-Smooth R code}

We store spatial data in a \texttt{tibble}, ordered by longitude first and
latitude next, with MLEs and observed information matrices stored as nested data in list columns. 
The following are the first few lines of the tibble \texttt{mle} used to store logistic regression parameters and information matrices:

{\footnotesize
\begin{verbatim}
mle
# # A tibble: 713 × 4
#    long   lat theta_hat J_hat
#   <dbl> <dbl> <list>    <list>
# 1   -15  36   <dbl [2]> <dbl [2 × 2]>
# 2   -15  37.5 <dbl [2]> <dbl [2 × 2]>
# 3   -15  39   <dbl [2]> <dbl [2 × 2]>
# # … with 710 more rows
\end{verbatim}
}

The R code below constructs the sparse precision matrix \texttt{Q}, the sparse information matrix \texttt{J\_hat}, and the vector of MLEs \texttt{theta\_hat}. 
The \texttt{solve} function provided by the \texttt{Matrix} package is used to solve the sparse linear system to calculate the smoothed parameter estimates.

{\footnotesize
\begin{verbatim}
library(Matrix)

## joint precision matrix (RW2D block matrix)
N_i = 23        # number of latitudes (rows)
N_j = 31        # number of longitudes (columns)
N   = N_i * N_j # number of grid points
kappa_hat = c(exp(8.73), exp(6.35)) # optimised hyperparameters
Di = bandSparse(n=N_i, k=0:1, sym=TRUE, 
                diag=list(c(1, rep(2, N_i-2), 1), rep(-1, N_i-1)))
Dj = bandSparse(n=N_j, k=0:1, sym=TRUE, 
                diag=list(c(1, rep(2, N_j-2), 1), rep(-1, N_j-1)))
D  = kronecker(Dj, Diagonal(N_i)) + kronecker(Diagonal(N_j), Di)
R  = crossprod(D)
Q  = bdiag(list(kappa_hat[1] * R, kappa_hat[2] * R))

## extract vector theta_hat and information block matrix from `mle`
theta_hat = do.call(c, mle$theta_hat)
J_hat     = bdiag(mle$J_hat)

## theta_hat and J_hat are ordered as (ababab...). we must reorder 
## them as (a...ab...b) to be compatible with Q
inds      = c(seq(1, 2*N, 2), seq(2, 2*N, 2))
theta_hat = theta_hat[inds]
J_hat     = J_hat[inds, inds]

## calculate Max-and-Smooth posterior mean 
theta_ms = drop(solve(Q + J_hat, J_hat %*% theta_hat))

## construct parameter vectors that align with the 
## longitudes and latitudes in `mle`
alpha_ms = theta_ms[1:N]
beta_ms  = theta_ms[1:N + N]
\end{verbatim}
}

\section{R-INLA smoothing code}

We start with a tibble containing latitude/longitude spatial data of MLEs \texttt{alpha\_hat} and their estimation error variances \texttt{alpha\_hat\_var}.

{\footnotesize
\begin{verbatim}
alpha_hat
## # A tibble: 713 × 4
##    long   lat alpha_hat alpha_hat_var
##   <dbl> <dbl>     <dbl>         <dbl>
## 1   -15  36       -6.65         15.9
## 2   -15  37.5     -3.28          1.43
## 3   -15  39       -2.43          1.04
## # … with 710 more rows

## add the spatial index s used in INLAs RW2D model. the function 
## as.numeric(factor(...)) assigns increasing integers starting at 1 to latitude
## and longitudes. 
N_i = 23
N_j = 31
alpha_hat = alpha_hat %>%
  mutate(i = as.numeric(factor(lat)), j = as.numeric(factor(long))) %>%
  mutate(s = (j - 1) * N_i + i)

## fit normal measurement error model with RW2D prior. fix error variances using
## `control.family` and `scale`
inla_alpha = inla(
  formula = alpha_hat ~ -1 + f(s, model='rw2d', nrow=N_i, ncol=N_j, constr=FALSE),
  data = alpha_hat, 
  family = 'gaussian', 
  control.family = list(hyper=list(prec=list(initial=0, fixed=TRUE))),
  scale = 1 / alpha_hat$alpha_hat_var
)

## extract posterior means of alpha
alpha_ms = inla_alpha$summary.random$s$mean
\end{verbatim}
}

The vector \texttt{alpha\_ms} contains the smoothed estimates of the intercept parameters in the same order as the MLEs given in the tibble \texttt{alpha\_hat}.

\section{CRPS and PIT analysis of NGR forecasts}

We repeat the leave-one-out verification of temperature forecasts postprocessed with NGR using the Continuous Ranked Probability Score \citep[CRPS,][]{matheson1976scoring} as the verification metric.
For a probabilistic forecast $f(x)$ issued as a Normal distribution with mean $\mu$ and variance $\sigma^2$, and verifying observation $y$, the CRPS is given by \citep{gneiting2005calibrated}
\begin{equation}
CRPS(f, y) = \sigma \left\{ z [2 \Phi(z) -1] + 2 \varphi(z) - \pi^{-1/2}\right\}
\end{equation}
where $z = (y - \mu)/\sigma$, and $\varphi(x)$ and $\Phi(x)$ are, respectively, the pdf and cdf of the standard normal distribution.

\begin{figure}
\begin{center}
\includegraphics{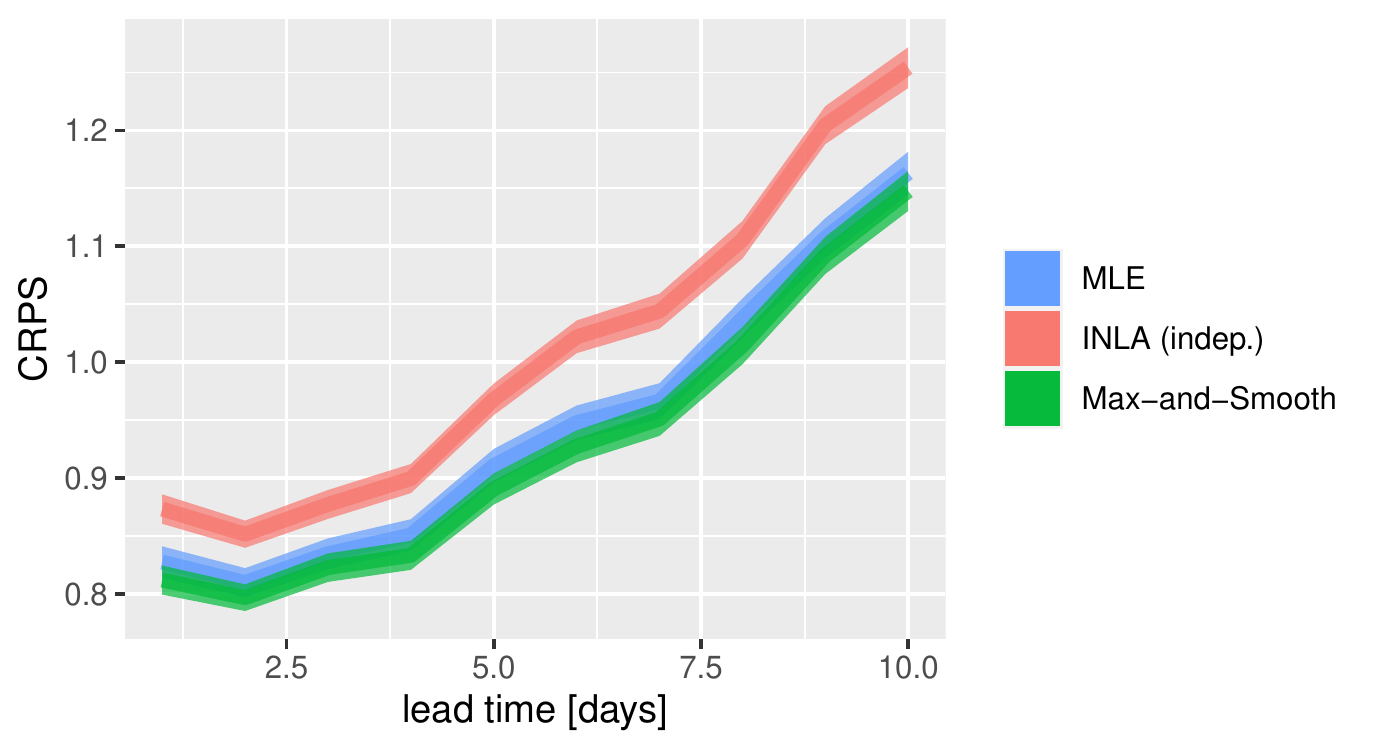}
\end{center}
\caption{Spatially averaged CRPS of NGR temperature forecasts. Smaller values indicate better forecasts.}
\label{fig:ngr-crps}
\end{figure}

Figure \ref{fig:ngr-crps} shows spatially averaged CRPS versus lead time similar to Fig.~\ref{fig:ngr-logscore} for Logscores.
In terms of CRPS the independently smoothed postprocessing parameters with R-INLA perform consistently worse than the the unsmoothed parameters.
When smoothing parameters jointly with Max-and-Smooth, forecasts improve consistently.

Histograms of Probability Integral Transforms (PITs) are used to assess forecast calibration \citep{gneiting2007probabilistic}.
A forecast at time $t$ and location $s$ is issued as a cumulative distribution function (cdf) $F_{s,t}(x)$, and verifying observation $y_{s,t}$ materialised at that instance.
The PIT value of that forecast is then $F_{s,t}(y_{s,t})$.
PIT values of a well-calibrated forecaster have a uniform distribution, and so the histogram of PIT values should be approximately flat.

\begin{figure}
\begin{center}
\includegraphics{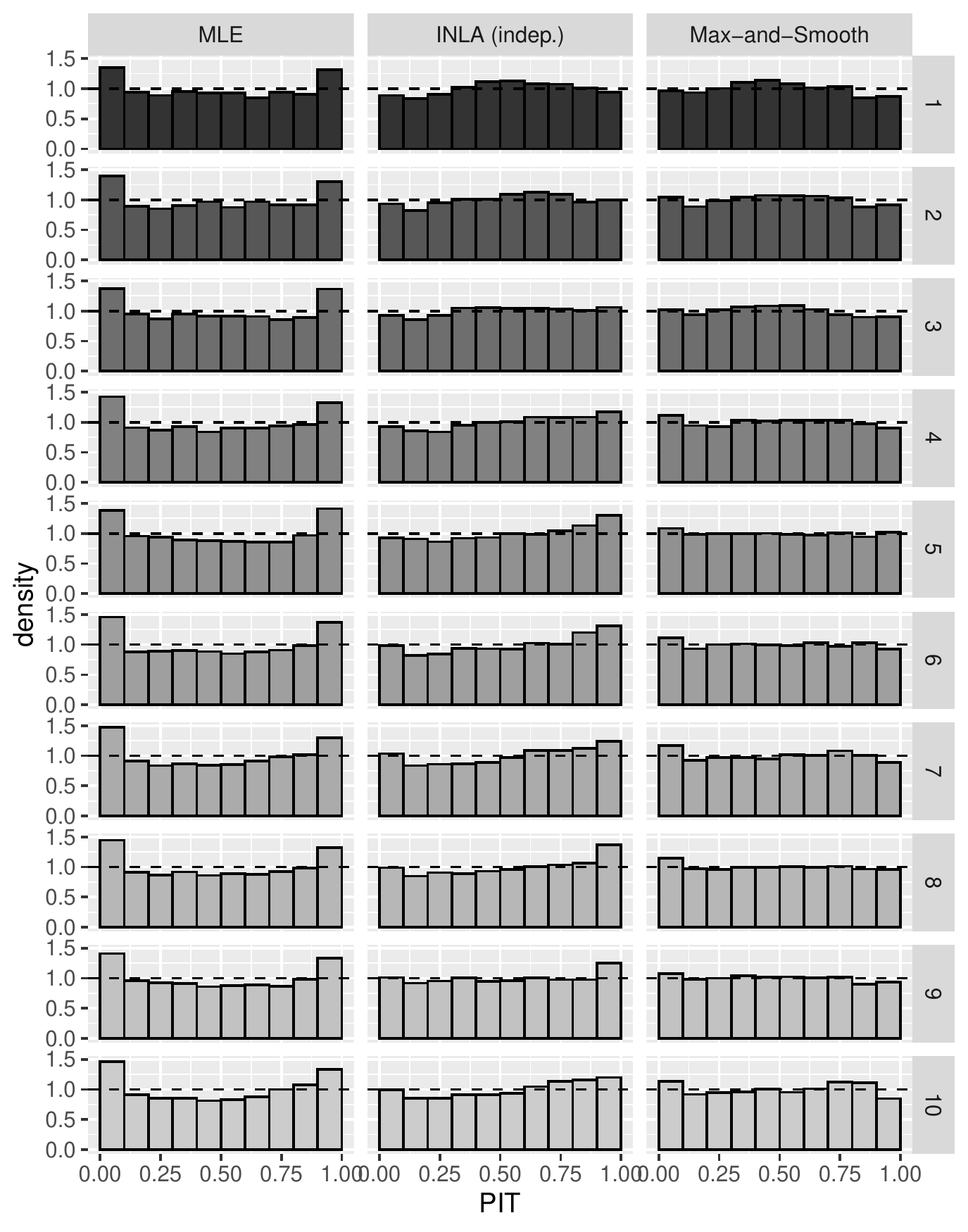}
\end{center}
\caption{PIT histograms of NGR temperature forecasts, separately for each lead time (rows). Well-calibrated forecasts have flat PIT histograms along the dashed horizontal line.}
\label{fig:ngr-pit}
\end{figure}

Figure~\ref{fig:ngr-pit} shows PIT histograms of (cross-validated) temperature forecasts postprocessed with NGR. 
We compare forecasts obtained with unsmoothed postprocessing parameters (MLE), forecasts obtained with independently smoothed parameters using R-INLA, and forecasts obtained with parameters smoothed with Max-and-Smooth.
Forecast distributions obtained with unsmoothed parameters are on average too narrow, as indicated by the overrepresented outer bins in the PIT histogram.
That effect is much reduced in the forecasts obtained with smoothed parameters.
The difference between independently and dependently smoothed parameters is less strong, but a case can be made that PIT histograms of Max-and-Smooth forecasts are more uniform than with indpendent smoothing, especially for lead times greater than 5 days.

\end{document}